\documentclass[12pt]{iopart}
\usepackage{graphicx}% Include figure files
\usepackage{dcolumn}% Align table columns on decimal point
\usepackage{bm}% bold math
\usepackage{hyperref}% Auto-generate hyperlinks
\renewcommand{\vec}[1]{{\mathbf{#1}}}

\newcommand{\beq}{\begin{eqnarray}}
\newcommand{\eeq}{\end{eqnarray}}
\renewcommand{\bs}{\bar\sigma}

\newcommand{\tn}{\tilde n}
\newcommand{\myfig}[3]

\begin{document}

\title{Mottness in High-Temperature Copper-Oxide Superconductors}

\author{Philip Phillips$^a$, Ting-Pong Choy and Robert G. Leigh}
\address{Department of Physics,
University of Illinois
1110 W. Green Street, Urbana, IL 61801, U.S.A}
\ead{$^a$dimer@uiuc.edu}
\begin{abstract}
The standard theory of metals, Fermi
liquid theory, hinges on the key assumption that although the
electrons
interact, the
low-energy excitation spectrum stands in a one-to-one
correspondence with that of a non-interacting system.  In the normal
state of the copper-oxide high-temperature superconductors, drastic
deviations from the Fermi liquid picture obtain, highlighted by a
pseudogap, broad spectral features and $T-$ linear resistivity.
A successful theory in this context must confront the highly
constraining scaling argument which
establishes that all 4-Fermi
interactions are irrelevant (except for pairing) at a Fermi surface. This argument lays plain that new low-energy degrees of
freedom are necessary.  This article focuses on the series of
experiments on the copper-oxide superconductors which reveal that the number of low-energy addition
states per electron per spin exceeds unity, in direct violation of the key
Fermi liquid tenet.  These experiments point to new degrees of freedom,
not made out of the elemental excitations, as the key mechanism by which
Fermi liquid theory breaks down in the cuprates.  A recent
theoretical advance which permits an explicit integration of the high
energy scale in the standard model for the cuprates reveals the source
of the new dynamical degrees of freedom at low energies, a charge 2e
bosonic field which has nothing to do with pairing but rather
represents the mixing with the high energy scales. We demonstrate
explicitly that at half-filling,
this new degree of freedom provides a dynamical mechanism for the
generation of the charge gap and antiferromagnetism in the insulating phase. 
  At finite doping, many of the anomalies of the
normal state of the cuprates including the pseudogap, $T-$linear
resistivity, and the mid-infrared band are reproduced.   A possible
route to superconductivity is explored.
\end{abstract}

\maketitle

\section{Introduction}

Superconductivity in the copper-oxide ceramics stands as a grand
challenge problem as its solution is fundamentally rooted in the
physics of strong coupling. In such problems, traditional
calculational schemes based on the properties of single free particles
fail.  Rather the physics of strong coupling resides in collective
behaviour, signified typically by the emergence of new degrees of
freedom at low energy. For example, in quantum-chromodynamics (QCD)
the propagating degrees of freedom in the infrared (IR) are bound states
not related straightforwardly to the ultra-violet (UV) scale
physics. The key perspective presented here is that similar physics stems
from the strong electron interactions in the copper-oxide
superconductors.  We will delineate precisely how the emergence of
collective behaviour at low energy accounts for many of the anomalous
properties of the normal state of the cuprates.

\begin{figure}
\centering
\includegraphics[width=6.0cm]{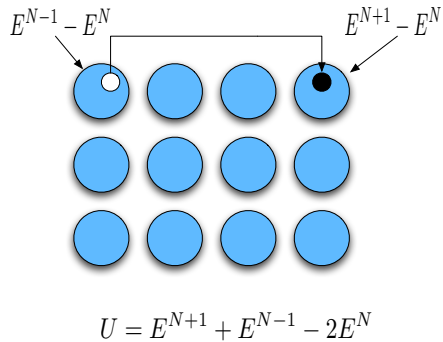}
\caption{A half-filled band as envisioned by Mott. Each blue circle
 represents a neutral atom with $N$ electrons and ground-state energy
 $E^{N}$.  The energy
 differences for electron removal and addition are explicitly shown.
Mott reasoned that no doubly occupied sites exist because at zero
temperature, $U=E^{N+1}+E^{N-1}-2E^{N}\gg 0$.  This is, of course, not
true. As a consequence the Mott gap must be thought of dynamically
rather than statically.}
\label{mott}
\end{figure}

That the cuprates embody strong coupling physics stems from the Mott
insulating\cite{mott} nature of the parent state.  Such
materials possess a half-filled band but insulate, nonetheless.  Their
insulating behaviour derives from the large on-site interaction two opposite-spin electrons
encounter whenever they doubly occupy the same lattice site.
For the cuprates\cite{stechel}, the
on-site electron repulsion is typically $U\simeq4eV$ whereas the
nearest-neighbour hopping matrix element is only $t\simeq0.4$eV. Although
double occupancy is costly, there is no symmetry principle that
forbids it even at
half-filling. In the original proposal by Mott\cite{mott} to explain why NiO
insulated, he assumed, as illustrated in Fig. (\ref{mott}), that each
Ni atom remained neutral because
\beq
U=E^{N+1}+E^{N-1}-2E^N
\eeq
dominates all other energy scales.  Here $E^{N}$ is the ground state
energy of an atom with $N$ valence electrons. For each Ni atom, $N=2$.
Hence, the zero-temperature
state envisioned by Mott is one in which no atom is excited with an
occupation of 
$N\pm 1$ electrons.  For NiO, this translates to no Ni$^{+++}$ or Ni$^{+}$ ions exist as
explicitly stated by Mott\cite{mott}.  However, it is well-known\cite{hubbard,harris,fulde} that the ground state of a
Mott insulator possesses doubly occupied sites at half-filling.  As a result, the simple cartoon\cite{mott} that the
Mott gap originates because double occupancy is forbidden is
incomplete. Some have advocated\cite{castellani} that in the Mott
insulator, doubly occupied sites are immobile whereas in the metal
they form a fluid.  This account requires an explicit dynamical
mechanism for the generation of the Mott gap.  However, the dynamical degrees of freedom leading
to the localization of
double occupancy have not been unearthed. We offer here an explicit
resolution of this problem.

\begin{figure}
\centering
\includegraphics[width=6.0cm]{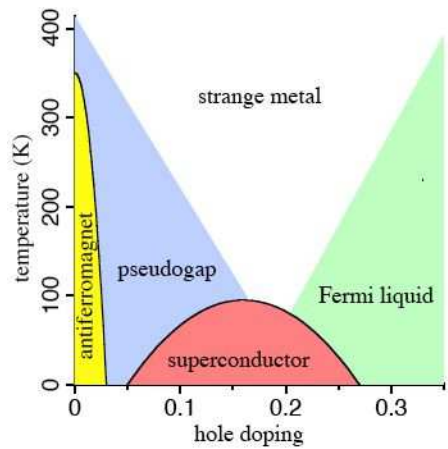}
\caption{Heuristic phase diagram of the copper-oxide
  superconductors.  In the
 strange metal, the resistivity is a linear function of
 temperature. In the pseudogap the single-particle density of states
 is suppressed without the onset of global phase coherence indicative
 of superconductivity.  As discussed in Section 3.6, the dome-shape of the superconducting region
 with an optimal doping level of $x_{\rm opt}\approx 0.17$ is not quantitatively
accurate.  See Fig. 19 for a more accurate determination of $x_{\rm opt}$.}
\label{pdiagram}
\end{figure}

A few of the properties of
doped Mott insulators are sketched in the phase diagram
in Fig. (\ref{pdiagram}).   Aside from
d$_{x^2-y^2}$ superconductivity, the
pseudogap, in which the single-particle density of states is
suppressed\cite{alloul,norman,timusk}, and the strange metal,
characterised by the ubiquitous $T-$linear
resistivity\cite{batlogg,ando}, stand out.  As the phase diagram suggests,
the pseudogap and strange metal phases are intimately related.  That
is, a correct theory of the pseudogap state of matter should at
higher temperatures yield a metallic phase in which the resistivity
scales as a linear function of temperature.  Nonetheless, numerous
proposals\cite{stripes,ddw,rvb,pfp,inco1,inco2,inco3} for the pseudogap abound that offer no resolution of
$T-$linear resistivity. Part of the problem is that a series of
associated phenomena, for example, incipient diamagnetism\cite{nernst}
indicative of
incoherent pairing\cite{inco1,inco2,inco3,inco4}, electronic
inhomogeneity\cite{stripes1,stripes2,stripes3,stripes4,stripes5,stripes}, time-reversal
symmetry breaking\cite{trsb1,trsb2,trsb3,tsrb4}, and quantum oscillations\cite{qoscill} in
the Hall conductivity,
possibly associated with the emergence of closed electron (not
hole) pockets in the first Brillouin zone (FBZ), obscure the
efficient cause of the pseudogap and its continuity with the strange metal.  Despite this
range of phenomena, a key experimental measure\cite{ando,raffy} of
the pseudogap onset is the temperature, $T^\ast$, at which the first
deviation from $T-$linear resistivity obtains.  As a consequence,
the physics underlying the strange metal must also yield a pseudogap
at lower temperatures.  Further, it must do so in a naturasl way. In
our work, we take the relationship between the strange metal and
pseudogap seriously and develop a theory\cite{ftm1,ftm2,ftm3} that explains both
simultaneously.  In addition, we show that the same theory is capable of
explaining other anomalies of the normal state such as   1) absence of
quasiparticles\cite{Kanigel} in the normal state, 2) the mid-infrared band in the
optical conductivity\cite{cooper,uchida1,opt0,opt1,opt2,opt3},
3) spectral weight transfer across the Mott gap, and 4) the high and
low-energy kinks in the electron removal spectrum.

While it has been acknowledged for some time\cite{polchinski} that the
normal state of the cuprates is incompatible with Fermi liquid theory, precisely what
replaces it has not been settled.  In a Fermi liquid, the
low-energy excitation spectrum stands in a one-to-one
correspondence with that of a non-interacting system. This
correspondence must clearly break down in the normal state of the
cuprates. The arguments of Polchinski\cite{polchinski} and others\cite{shankar,others,others2} make it clear
that breaking Fermi liquid theory in $d=2$ requires new degrees of freedom at
low energy, not simply 4-fermion interactions as they are all (except
for pairing) irrelevant at the Fermi liquid fixed point.  One possible
origin of the new degrees of freedom\cite{ruckenstein} is if spectral weight transfer between high and low
energies mediates new electronic states at low energy.  As a result,
new states will emerge at low energy that have no counterpart in the
non-interacting system.  We show quite generally that this state of
affairs obtains in the minimal model for a doped Mott insulator,
namely the Hubbard model.  Refinements of this model to include more details of the copper-oxide
plane also retain this feature.  We establish this result first
through a
simple physical argument which lays plain that in a doped Mott
insulator, the phase space available for adding a particle exceeds the
number of ways electrons can be added at low energy.  Consequently,
some new degrees of freedom not made out of the elemental excitations
must reside in the low-energy spectrum.   By explicitly integrating
out the degrees of freedom far away from the chemical potential, the
Wilsonian program for constructing a proper low-energy theory, we show that this new
excitation is a charge 2e bosonic field that in no way has anything to
do with pairing.  It is from this new degree of freedom that the
pseudogap and T-linear resistivity follow immediately. Since this
physics arises without any appeal to some further fact but relies only
on the 
strong correlations of the doped Mott state, we have successfully
isolated the efficient cause of the pseudogap.  The associated phenomena mentioned
above are supervenient on rather than central to the physics of the
normal state.   This review is
organised as follows.  In the next section, we discuss the
experimental evidence for spectral weight transfer and show that it
requires new degrees of freedom at low energy not made out of the
elemental excitations. In section II, we derive the exact low-energy
theory by formally integrating out the degrees of freedom far away
from the chemical potential.  In Section III, we compare the
predictions of the theory with experiment.  We close with a
perspective on the remaining problem of superconductivity.

\section{Mottness}

The origin of the Mott insulating state is subtle for two related reasons.  First, the Mott gap cannot be easily deduced from the bare degrees of freedom in
a model Hamiltonian.   As remarked in the introduction, even in the Hubbard
model, the ground state contains admixtures with the degrees of freedom, namely double occupancy, that lie above the gap.  That is, if one were to write the bare electron operator\cite{hubbard} 
\beq\label{sep}
c_{i\sigma}=(1-n_{i-\sigma})c_{i\sigma}+n_{i-\sigma}c_{i\sigma}
\eeq
as a sum of two operators, one of which vanishes on doubly occupied sites, $\xi_{i\sigma}=(1-n_{i-\sigma})c_{i\sigma}$ and its complement which is only non-zero when a site is doubly occupied, $\eta_{i\sigma}=n_{i-\sigma}c_{i\sigma}$,
one would see immediately that such a separation is not canonical.  As
a result, $\eta_{i\sigma}$ and $\xi_{i\sigma}$ have a non-zero overlap
and hence they do ${\bf not}$ propagate independently as would be
required for them to be gapped.  In fact, it is unclear precisely how
to write down a set of canonically defined fermionic operators that do
become gapped as a result of the energy cost for double occupancy.
This is the Mott problem.  Its persistence has led
Laughlin\cite{laughlin} to
assert that the Mott problem is fictitious and, in reality, does not
exist.   As mentioned in the preceding
section, the real problem is that the Mott gap is fundamentally
dynamical in nature.  That it is difficult to write down the
precise degrees of freedom that are becoming gapped is just a symptom of
this fact.   As will become clear from this review, the dynamical
degrees of freedom that ultimately produce the Mott gap only appear
when the high-energy scale is integrated out exactly. 

Second, all known Mott insulators order antiferromagnetically at sufficiently low temperature. To illustrate, two electrons on neighbouring sites with opposite spins can
exchange their spins.
This process proceeds through an intermediate state in which one of
the sites
is doubly occupied and hence the corresponding matrix element scales as $t^2/U$. Antiferromagnetism in
the cuprates arises from this mechanism.  This mechanism
is distinct from the weak-coupling Slater\cite{slater} process in which a
half-filled band orders as a result of nesting at
$Q=(\pi,\pi)$. While antiferromagnetism is certainly part of the Mott
insulating story, it leaves much unexplained.  It is that explanatory residue, namely the properties of Mott
insulators which do not necessitate ordering, we refer to as
{\it Mottness}.  A simple property unexplained by ordering is the Mott gap itself.  Above any
temperature associated with ordering, an optical gap
obtains\cite{cooper,uchida1,linio}.  Another such property is spectral weight
transfer.

\subsection{Spectral Weight Transfer}

\begin{figure}
\centering
\includegraphics[width=6.0cm]{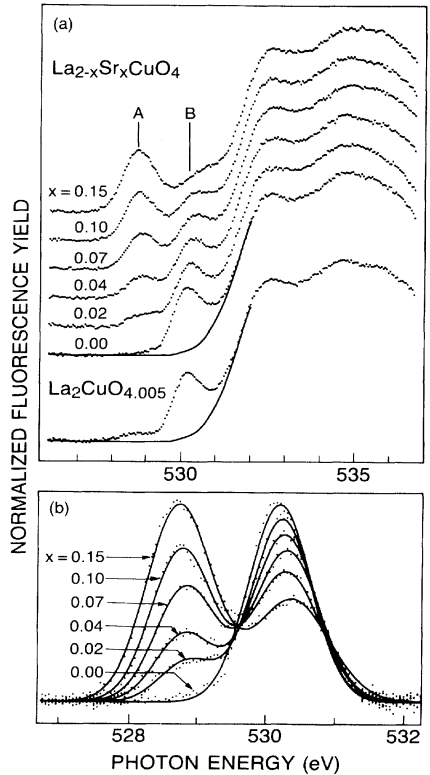}
\caption{a) Normalized flourescence\cite{chen} yield at the O K edge of
 La$_{2-x}$Sr$_x$CuO$_{4+\delta}$.  In the undoped sample, the only
 absorption occurs at 530eV, indicated by B.  Upon doping the
 intensity at B is transferred to the feature at A, located at
 528eV.  b) Gaussian fits to the absorption features at A and B with
 the background subtracted. Reprinted from Chen, et al. Phys. Rev. Lett. {\bf 66}, 104 (1991).}
\label{fy}
\end{figure}

While what constitutes the minimal model for the cuprates can
certainly be debated, it is clear\cite{p2,p3,p4} that regardless of the model, the
largest energy scale arises from doubly occupying the copper
$d_{x^2-y^2}$ orbital.  This orbital can hybridise with the in-plane
$p_x$ and $p_y$ orbitals and hence a two-band model is natural.  Since
our emphasis is on the interplay between the high and low-energy
scales, we simplify to a one-band Hubbard\cite{hubbard} model
\beq\label{hubb}
H_{\rm Hubb}&=&-t\sum_{i,j,\sigma} g_{ij} c^\dagger_{i,\sigma}c_{j,\sigma}+U\sum_{i,\sigma} c^\dagger_{i,\uparrow}c^\dagger_{i,\downarrow}c_{i,\downarrow}c_{i,\uparrow},
\eeq
where $i,j$ label lattice sites, $g_{ij}$ is equal to one iff $i,j$
are nearest neighbours, $c_{i\sigma}$ annihilates an electron with
spin $\sigma$ on lattice site $i$, $t$ is the nearest-neighbour
hopping matrix element and $U$ the energy cost when two electrons
doubly occupy the same site. Our conclusions carry over naturally to
any n-band model of the cuprates as long as the largest energy scale is
the on-site energy, $U$ in Eq. (\ref{hubb}).  That the dynamics of the charge
carriers in the cuprates are captured by this model was confirmed
by Oxygen 1s x-ray absorption\cite{chen} on La$_{2-x}$Sr$_x$CuO$_4$.  In such
experiments, an electron is promoted from the core 1s to an
unoccupied level.  The experimental observable is the fluorescence
yield as a function of energy as electrons relax back to the
valence states.  The experiments, Fig. (\ref{fy}), show that at
$x=0$, all the available states lie at 530eV.  As a function of
doping, the intensity in the high-energy peak decreases and is
transferred to states at 528eV.  In fact, the lower peak grows faster
than $2x$ while the upper peak decreases faster than $1-x$.  The separation between these two
peaks is the optical gap in the parent insulating material. Though this
observation of transfer of spectral weight from high to low energy is
not expected in a semiconductor or a band insulator, it is certainly
not an anomaly in strongly correlated systems.  In fact, it is the
fingerprint of Mottness as it has been observed in the classic
Mott system NiO upon Li doping\cite{linio} and in all optical
conductivity
measurements on the cuprates\cite{cooper,uchida1,opt1,opt2,opt3} above
any temperature having to do with ordering.

\begin{figure}
\centering
\includegraphics[width=9.0cm]{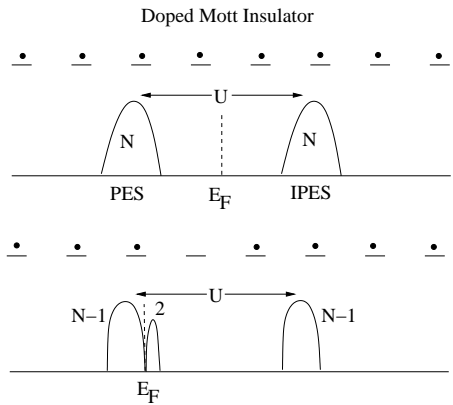}
\caption{Evolution of the single-particle density of states from
 half-filling to the one-hole limit in a doped Mott insulator
 described by the Hubbard model.  Removal of an electron results
 in two empty states at low energy as opposed to one in the
 band-insulator limit. The key difference with the Fermi liquid is
 that the total weight spectral weight carried by the lower Hubbard
 band (analogue of the valence band in a Fermi liquid) is not a
 constant but a function of the filling. }
\label{spectrans}
\end{figure}

This generic behaviour of spectral weight transfer is captured by the
Hubbard model. To illustrate, consider the half-filled Hubbard model.
A charge gap splits the spectrum into two parts, lower and upper
Hubbard bands.  Roughly, the lower Hubbard band LHB) describes particle
motion on empty sites while particle motion on already singly occupied
sites is captured by the upper Hubbard band (UHB). This relationship is only
approximate because the UHB and LHB are mixed so that there are states
in the LHB that have some doubly occupied character.  To understand
spectral weight transfer, we start in the atomic limit in which there
is a clean gap of order $U$ between the UHB  and LHB.  For a system
containing N electrons on N sites, the weight of the LHB is N
corresponding to N ways to remove an electron.  The corresponding
weight in the UHB is N as well as there are N ways to add
an electron to the system.  These bands are
shown in Fig. (\ref{spectrans}).  Consequently, adding a hole in the
atomic limit 
decreases the electron removal spectrum in the LHB by one state.  The weight in the UHB
is also affected as there are now N-1 ways to create a doubly occupied
site.  This leaves two states unaccounted for.  The two extra states are part of the addition spectrum at low
energies and correspond to the two ways of occupying the empty site by
either a spin up or a spin down electron.  In the atomic limit, the
number of addition states scales as $2x$\cite{sawatzky,stechel2} when $x$ holes are created.
In a semiconductor or a Fermi liquid, the number of addition states
would be strictly $x$.  Experimentally\cite{chen,linio,cooper,uchida1,opt1,opt2,opt3}, however, the low-energy
spectral weight (LESW) grows faster than $2x$.   The excess of $2x$ can be
understood simply by turning on the hopping\cite{harris}.  When the hopping is
non-zero, empty sites are created as a result of the creation of
double occupancy.  Such events increase the number of
available states for particle addition and as a consequence the LESW increases faster
than $2x$.  It is important to recall that the argument leading to the
LESW exceeding $2x$ relies on the strong coupling limit.  If this
limit is not relevant to the ground state at a particular filling, the
previous argument fails.

\subsection{Breakdown of Fermi Liquid Theory: More than just Electrons}

A natural question arises.  Is spectral weight transfer important?
A way of gauging importance is to determine if spectral weight
transfer plays any role in a low-energy theory.   A low-energy theory
is properly considered to be natural if there are no relevant
perturbations.  Several years ago, Polchinski\cite{polchinski} and others\cite{shankar,others,others2}
considered Fermi liquid theory from the standpoint of
renormalisation.  They found\cite{polchinski,shankar,others,others2} that as long as one posits that the charge carriers are
electrons, there are no relevant interactions (except for pairing)
that destroy the Fermi liquid state.  The setup\cite{polchinski} is as follows.  Decompose the momenta into the Fermi momentum
and a component orthogonal to the Fermi surface
\beq
\vec{p} = \vec{k} + \vec{\ell}.
\eeq
Here $\vec{l}$ is the component orthogonal to the Fermi surface.
Then consider scaling of energy and momentum towards the Fermi surface, in other words
\beq
E \to uE, \quad \vec{k} \to \vec{k}, \quad \vec{\ell} \to u \vec{\ell},
\eeq
where $u$ is the scaling parameter.
To quadratic order, the action is
\beq
S = \int dt d^3 \vec{p} [ i \psi^{*} (\vec{p}) \partial_t \psi (\vec{p})-
(E(\vec{p}) -E_F(\vec{p}))\psi^{*} (\vec{p})  \psi (\vec{p})].\nonumber
\eeq
Hence, close to the Fermi surface
\beq
E(\vec{p}) -E_F(\vec{p}) \sim \ell v_f, \quad v_F= \partial_{\vec{p}} E
\eeq
so that after scaling towards the Fermi surface (note that also
$t \to u^{-1} t$) one finds that
\beq
\psi \to u^{-1/2} \psi.
\eeq
Consider now the four-fermion interaction.  The argument to show that
such interactions are irrelevant is particularly simple.  In terms of
powers of the scaling parameter, $u$, the measure
over time contributes one negative power,
the measure over the momenta orthogonal to the Fermi surface 4 powers and the
4-fermi interaction $4/2$ negative powers.
The delta function over the 4-momenta generically does not scale.
Hence, the overall scaling of the four-Fermi interaction is governed by
$u^{-1+4-4/2} = u^{1}$
and hence is irrelevant as the power of $u$ is positive.
The only exception to this argument if inversion symmetry  is present
is the Cooper pairing interaction.  Consequently, as long as the
charge carriers carry unit charge, there are no relevant interactions
that destroy Fermi liquid theory.   In the context of the
cuprates, this argument is particularly powerful as it implies that in order to
explain $T-$linear resistivity, some new emergent degrees of freedom
that have nothing to do with the electrons must be present.   There have
been attempts to circumvent this argument in the literature that
amount to essentially free field theory. In light of the above
argument, such attempts must reduce to Fermi liquid theory and hence
must yield $T^2$ resistivity.  Others\cite{leclair1,leclair2} have directly confronted
the Polchinski\cite{polchinski} argument and added extra derivative
couplings to the Fermi liquid action.  However, the relationship of
such continuum models\cite{leclair1,leclair2} to any concrete realisation of Mott physics is
not clear.

\begin{figure}
\centering
\includegraphics[width=3.0cm,angle=-90.0]{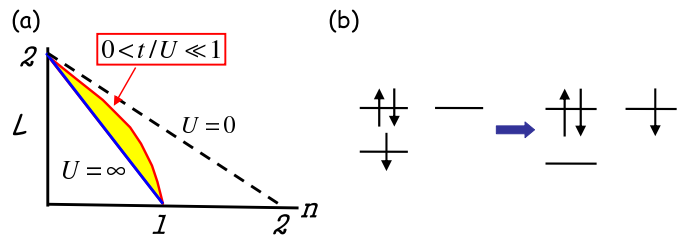}
\caption{a) Integrated low-energy spectral
weight, $L$, defined in Eq. (\ref{dos}), as a function of the
electron filling, n: 1) the dashed line is the non-interacting limit, vanishing on-site
interaction ($U=0$), in which $L=2-n$, 2) atomic limit
(blue line) of a doped
Mott insulator, $U=\infty$, in which $L=2(1-n)=2x$, $x$ the doping
level and 3) a real Mott insulator in which $0<t/U\ll 1$, red curve. For
$0<t/U\ll 1$, $L$ must lie strictly above the $U=\infty$ limit and hence $L>2x$ away from the
atomic limit. (b) Hopping processes mediated by the $t/U$ terms in
the expansion of the projected transformed operators in terms of the bare electron operators (see Eq. (\ref{trans})). As a
result of the $t/U$ terms in Eq. (\ref{trans}), the low-energy
theory in terms of the bare fermions does not preserve double
occupancy. The process
shown here illustrates that mixing between the high and low-energy
scales obtains only if double occupancy neighbours a hole. In the exact low-energy theory, such processes are mediated by the new
degree of freedom, $\varphi_i$, the charge $2e$ bosonic field which
binds a hole and produces a new charge $e$ excitation, the collective
excitation in a doped Mott insulator.}
\label{dos}
\end{figure}

It is straightforward to show that dynamical spectral weight transfer in a doped
Mott insulator leads to a breakdown of the Fermi liquid
picture and the emergence of new low-energy degrees of freedom. The
interactions of the electrons with the
new degrees of freedom can be formulated as a natural theory in which
the electron spin-spin interaction is sub-dominant.   As will be seen,
the interactions with the new degrees of freedom govern all the
physics that is independent of ordering.  In this sense, we arrive at
a natural separation between spin-ordering and Mott physics.  To
proceed, we define the number of single-particle
addition states per site at low energy,
\beq\label{L}
L=\int_\mu^\Lambda N(\omega)d\omega,
\eeq
as the integral of the single-particle density of states
($N(\omega)$) from the chemical potential, $\mu$, to a cutoff energy
scale, $\Lambda$, demarcating the IR and UV scales. As long as
$\Lambda$ is chosen to exclude the high-energy scale, $L$ is a
well-defined quantity which simply counts the number of states in the
unoccupied part of the spectrum at low energy.  We compare this
quantity to the number of ways an electron can be added to
the holes created by the dopants. We call this quantity $n_h$.  Our usage of `ways' here refers to the spin degree of freedom of the electron only and not to combinatorics.   From the perspective of
single-particle physics, the intensity of a band is always equal to
the number of electrons the band can hold. Hence, strict adherence to
the single-particle picture requires that $L=n_h$, implying
that the number of low-energy addition states per electron per spin is
identically unity.  For example, as shown in Fig. (\ref{dos}), in a
non-interacting system, $L=2-n=n_h$.  The same is true for a Fermi
liquid as can be seen from the fact that
\beq
\int_{-\infty}^{\epsilon_F} N(\omega)d\omega=n.
\eeq
Since the integral over all energies must yield $2$, it follows that
$L=2-n$.
Hence, strictly for a Fermi liquid, $L/n_h=1$ as is dictated by the
basic Landau tenet that the number of bare electrons at a given
chemical potential equals the number of Fermi excitations
(quasiparticles) in the interacting system.

\begin{figure}
\centering
\includegraphics[width=8.0cm]{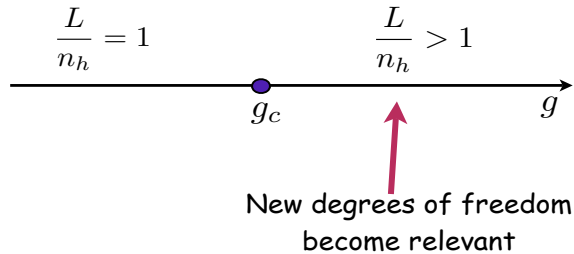}
\caption{Schematic depiction of electronic models based on the
 magnitude of $L/n_h$. In a Fermi liquid, $L/n_h=1$.  $L/n_h>1$ necessarily
leads to a break down of the Fermi liquid picture as new degrees of
freedom not made out of the electrons are needed.  $L/n_h>1$ appears
to be the generic way in which Fermi liquid theory breaks down in a doped
Mott insulator.}
\label{nfl}
\end{figure}

A doped Mott system is quite different because the total spectral
weight in the lower-Hubbard band is not simply $2$ but rather
determined by the electron filling.  Consider a Mott system in the
atomic limit.  As shown in the previous section, $L=2x$ is the exact result in
the atomic limit because creating a hole leaves behind an empty site which
can be occupied by either a spin-up or a spin-down electron.  Likewise,
the number of ways electrons  can occupy the empty sites is $n_h=2x$.
Hence, even in the atomic limit of a doped Mott insulator, $L/n_h=1$.  However, real Mott systems are not in the
atomic limit.  In strong coupling, finite hopping with matrix element $t$ creates double
occupancy, and as a result empty sites with weight $t/U$. Such empty
sites with fractional weight contribute to $L$ as shown first by
Harris and Lange\cite{harris}.   In fact, every order in perturbation theory
contributes to $L$.  Consequently,
when $0<t/U\ll 1$, $L$ is strictly larger than $2x$.  Such hopping
processes
or quantum fluctuations
do not affect the number of electrons that can be added to the system,
however.  Since $n_h$ remains fixed at $2x$, in a real doped Mott system, $L/n_h>1$.  Consequently, in contrast to a Fermi liquid, simply counting
the number of electrons that can be added does {\it not} exhaust the
available phase space to add a particle at low energy. That is,
addition states that do not have the quantum numbers of an electron
must exist as illustrated in Fig. (\ref{nfl}).  Since the number of
ways of adding a particle exceeds the number of electrons that can be
added, the additional states must be gapped to the addition of an
electron.  This gap can manifest itself straightforwardly as a
depression in the density of states at the chemical potential or more
subtely as a reconstruction\cite{drew} of the non-interacting Fermi surface, for
example, one that has electron (or hole) pockets that shrink in size
as the doping decreases. Numerical simulations show\cite{konstanzeros}
that such reconstructions do not necessitate broken symmetry but obtain
entirely from the strong correlations in a doped Mott insulator.  In
either case, the one-to-one correspondence between the excitation
spectrum in the free and interacting systems breaks down.  In doped
Mott systems, this breakdown arises entirely from spectral weight transfer. While it has
been known for some time\cite{harris} that $L>2x$ at strong coupling in a doped
Mott insulator based on the Hubbard model, that this simple fact
implies a pseudogap (whose dimensional dependence is discussed in
Section 3.2) has
not been deduced previously. Thus,
additional degrees of freedom at low-energy, not made out of the elemental
excitations, emerge in a low-energy reduction of a doped Mott
insulator at strong coupling.  Note, it is only the spectral weight in excess of $2x$
that creates the new physics.  Hence, the physics governed by the new
degrees of freedom has nothing to do with gauge\cite{gauge1,gauge2,gauge3} fields that engineer the no double occupancy constraint in standard treatments of
doped Mott insulators.  
The physics referred to here is precisely the part thrown away in such
treatments.  We refer to this contribution as the
dynamical part of the spectral weight. This dynamical part of $L$
arises through exchanges with doubly occupied sites. Hence, although
$t/U$ is approximately $1/10$ in the cuprates, the $t/U$ corrections
to $L$ must be retained as they mediate fundamentally new physics that
is quantifiable.   Namely, it is through these corrections that a Fermi
liquid description breaks down as depicted in Fig. (\ref{nfl}).  As a result, one
might imagine that the new physics is associated with some new collective
excitation with charge $2e$.  The new theory we construct, in which we
integrate out exactly the high energy scale, has such a degree of
freedom which does in fact mediate the dynamical part of $L$.  In
light of Polchinski's argument\cite{polchinski}, any non-Fermi liquid
behaviour must emerge from the new collective charge $2e$ excitation.  
We show that this is in fact the case.

\subsection{Perturbative Approaches}

It is possible to account for the dynamical contribution to the
low-energy spectral weight using degenerate perturbation theory.  While
this method does not
shed any light on the missing degree of freedom at low energy, it does
serve to illustrate that the limits of $U\rightarrow\infty$ and the
thermodynamic limit, $N\rightarrow\infty$, do not commute. In fact, it is this
lack of commutativity that gives rise to $L>2x$.  We review this
method here as it does serve to motivate our eventual analysis.

The goal of perturbative approaches\cite{eskes,slavery,spalek,girvin,sasha,anderson} in this context is to bring the
Hubbard model into diagonal form with respect to double occupancy. As
with any matrix diagonalization problem, the new basis which makes
double occupancy a good quantum number involves some linear
combination of the old states.  The subtlety that this introduces is that the no
double-occupancy condition applies only to the transformed fermions
{\bf not} to the original bare electrons.  This is an oft-overlooked fact that
has led to much confusion over what precisely the accepted low-energy
reduction of the Hubbard model, namely the t-J model\cite{rice}, entails.   We
review the derivation with an eye on isolating the processes which
lead to dynamical spectral weight transfer.  Let $f_{i\sigma}$ be the
dressed operators which make the Hubbard model block diagonal with
respect to ``double occupancy.''  Following Eskes et al.\cite{eskes},
for any operator $O$, we define  $\tilde O$
such that $O\equiv {\bf O}(c)$ and $\tilde{O}\equiv {\bf O}(f)$,
simply by replacing the Fermi operators $c_{i\sigma}$ with the
transformed fermions $f_{i\sigma}$.  To block diagonalise the Hubbard
model,
\beq
H[f]\equiv e^{S[f]}\tilde H[f] e^{-S[f]},
\eeq
one constructs a similarity transformation $S[f]$ which connects
sectors that differ by at most one `fictive' doubly occupied
site, that is, a doubly occupied site in the transformed basis. To lowest order,
\beq
S^{(1)}=\frac{1}{U} \left( \tilde T_{+1} - \tilde T_{-1}\right).
\eeq
where
\beq
\tilde T_{+1}=-t\sum_{i,j,\sigma} g_{ij} \tilde n_{i\bar\sigma} f^\dagger_{i\sigma} f_{j\sigma} (1-\tilde n_{j\bar\sigma}),
\eeq
which increases the quantum number $\tilde V =\sum_i \tilde n_{i\uparrow} \tilde n_{i\downarrow}$ by one. Likewise, $\tilde T_{-1} = (T_{+1})^\dagger$ decreases $\tilde V$ by one. In the new basis,
$[H,\tilde V]=0$,
implying that double occupation of the transformed fermions
is a good quantum number, and all of the eigenstates
can be indexed as such. This does not mean that $[H,V]=0$. If it
were, there would have been no reason to do the similarity transformation in
the first place. $\tilde V$, and
not $V$, is conserved. Assuming that $V$ is the conserved
quantity results in a spurious local SU(2)\cite{lsu21,lsu22}
symmetry in the strong-coupling limit at
half-filling.

To expose the dynamical contribution to spectral weight transfer,
we focus on the relationship between the physical and transformed
fermions.  As expected in any degenerate perturbation scheme, the bare
fermions,
\beq\label{trans1}
c_{i\sigma}&=&e^Sf_{i\sigma}e^{-S}
\simeq f_{i\sigma}-\frac{t}{U}\sum_{\langle j, i\rangle}
\left[(\tn_{j\bs}-\tn_{i\bs})f_{j\sigma}
-f^\dagger_{j\bs}f_{i\sigma}f_{i\bs}+f^\dagger_{i\bs}f_{i\sigma}f_{j\bs}\right],
\eeq
are linear combinations of the multiparticle states in the transformed
basis. We invert this relationship to find that
\beq
f_{i\sigma}\simeq c_{i\sigma}+\frac{t}{U}\sum_{j} g_{ij}X_{ij\sigma}
\eeq
where
\beq
X_{ij\sigma}=\left[(n_{j\bs}-n_{i\bs})c_{j\sigma}-c^\dagger_{j\bs}c_{i\sigma}c_{i\bs}+c^\dagger_{i\bs}c_{i\sigma}c_{j\bs}\right].
\eeq
Since the low energy theory is captured by the sector in the transformed basis
which has no double occupancy, it is most relevant to focus on the
form of the projected transformed fermions. Using the relations above,
we find that as expected, the projected transformed fermions
\beq\label{trans}
(1-\tn_{i\bs})f_{i\sigma}&\simeq &(1-n_{i\bs})c_{i\sigma}+\frac{t}{U}V_\sigma
c_{i\bs}^\dagger b_i+\frac{t}{U}\sum_{j}g_{ij}\left[
n_{j\bs}c_{j\sigma}+n_{i\bs}(1-n_{j\bs})c_{j\sigma}\right.\nonumber\\
&&\left.+(1-n_{j\bs})\left(c_{j\sigma}^\dagger
c_{i\sigma}-c_{j\sigma}c^\dagger_{i\sigma}\right)c_{i\bs}\right]
\eeq
involve double occupancy in the bare fermion basis. 
Here $V_\sigma=-V_{\bar\sigma}=1$ and
$b_i= \sum_{j\sigma} V_\sigma c_{i\sigma}c_{j\bs}$ where $j$ is summed over the nearest neighbors of $i$. The projected
bare fermion, $(1-n_{i\bs})c_{i\sigma}$, yields the $2x$ sum
rule, whereas it is the admixture with the doubly occupied sector
that mediates the $t/U$ corrections.  A process mediated by these
terms is shown in Fig. (\ref{dos}).  This can be seen more clearly by
computing $L$ directly using Eq. (\ref{trans1}). The standard
treatment\cite{gauge1,gauge2,gauge3,rice} of the t-J model ignores the dynamical corrections as a hard
projection scheme is implemented in which the no double occupancy
condition applies not only to the transformed but also to the bare
fermions.  As we have pointed out in the introduction, the physics left
out by projecting out double occupancy is important because it tells
us immediately that $L/n_h>1$ as can be seen from the expression for
$L$:
%%\begin{widetext}
\beq
L &\equiv& 2 \langle (1-n_{i\uparrow}) (1-n_{i\downarrow}) \rangle\\
&=& 2 \langle (1-\tilde n_{i\uparrow}) (1-\tilde n_{i\downarrow}) \rangle
+ \frac{2t}{U}\sum_{i,j,\sigma} g_{ij}\langle f_{i\sigma}^\dagger
\left[(\tn_{j\bs}-\tn_{i\bs})f_{j\sigma} f^\dagger_{j\bs}f_{i\sigma}f_{i\bs}\right.\nonumber\\
&+&\left. f^\dagger_{i\bs}f_{i\sigma}f_{j\bs}\right]
(1-\tilde n_{i\bar\sigma} ) + h.c.\rangle
\eeq
%%\end{widetext}
As is evident, $2 \langle (1-\tilde n_{i\uparrow}) (1-\tilde n_{i\downarrow}) \rangle = 2x$ in the projected Hilbert space of the dressed fermions
which corresponds to the $2x$ sum rule of the static part in the low
energy spectral weight. The dynamical part of $L$ arises from the
$t/U$ corrections.   It is these corrections that prevent the operator
in
Eq. (\ref{trans}) from being regarded as a free excitation. Rather it
describes a non-Fermi liquid ($L/n_h>1$).

However, it is particularly cumbersome to extract physical insight
from the canonical transformation method.  The primary reason is that
any information regarding the bare fermions requires that the
similarity transformation be undone when any experimentally relevant
quantity is calculated.  Consider for example the electron spectral
function. In the hard projected version of the $t-J$ 
model\cite{gauge1,gauge2,gauge3,rice}, the
electron spectral function is assumed to be given by simply the
time-ordered anticommutator of the transformed fermions. However,
Eq. (\ref{trans1}) illustrates that this is not so.  In actuality, the
single-particle Green function,
%%\begin{widetext}
\beq
\label{eq:green-c}
G(\vec k,\omega) &=& -i FT \langle Tc_{i\sigma} c_{j\sigma}^\dagger \rangle= -i FT \langle T f_{i\sigma} f_{j\sigma}^\dagger \rangle\nonumber\\
&+&
i FT \frac{t}{U} \sum_k g_{ik}\langle T \left[(\tn_{k\bs}-\tn_{i\bs})f_{k\sigma} f^\dagger_{k\bs}f_{i\sigma}f_{i\bs}+f^\dagger_{i\bs}f_{i\sigma}f_{k\bs}\right] f_{j\sigma}^\dagger \rangle\nonumber\\
&&+
i FT \frac{t}{U} \sum_k g_{jk}\langle T f_{i\sigma}
\left[(\tn_{k\bs}-\tn_{j\bs})f_{k\sigma}
 f^\dagger_{k\bs}f_{j\sigma}f_{j\bs}+f^\dagger_{j\bs}f_{j\sigma}f_{k\bs}\right]^\dagger \rangle,\nonumber\\
\eeq
%%\end{widetext}
has
$t/U$ corrections in the transformed basis.  Although these corrections are naively down by a
factor of $t/U$ relative to the projected part, their contribution
cannot be ignored because it is in these corrections that the explicit
non-Fermi liquid behaviour is hidden.  To illustrate, all calculations
on the $t-J$ model of the single hole
problem\cite{shole1,shole2,shole22,shole3} in a quantum antiferromagnet yield
a finite value of Z proportional to $J/t$. However, in the Hubbard
model, the situation is not as clear.  Simulations on finite-size systems reveal that $Z\propto L^{-\alpha}$ where $\alpha>0$ and hence tends to zero as the system size increases.  While this calculaton is not conclusive, it is consistent with the fact that similar dynamical mean-field
treatments of the t-J and Hubbard models at finite doping yield a finite conductivity
as $T\rightarrow 0$ in the $t-J$ model\cite{rosch,prelovsek} but a vanishing value in the
Hubbard model as $T\rightarrow 0$\cite{holeloc}.  These differences are summarized
in Table (\ref{tjsumm}). The most striking results in Table
(\ref{tjsumm}) are those for the exponents governing the asymptotic fall-off
of the density correlations, $\alpha_c$, and momentum distribution
functions ($\theta$) in the $t-J$ (with $t=J$, the supersymmetric
point) and Hubbard models in $d=1$.  Here these quantities can be
obtained exactly\cite{yang2,yang,korepin} for both models using Bethe
ansatz.  In the d=1 Hubbard model, the exponent $\theta$ approaches\cite{yang,korepin}
$1/8$ asympotically as $U\rightarrow\infty$ for any filling.  By
contrast in the $t-J$ model\cite{yang2}, $\theta$ is strongly dependent on doping
with a value of $1/8$ at half-filling and vanishing to zero as $n$
decreases.  More surprising, the exponent $\alpha_c$ remains pinned\cite{yang2,yang} at
$2$ for the $U\rightarrow\infty$ limit of the Hubbard model at any
filling.  In fact, at any value of $U$, $\alpha_c=2$\cite{yang2,yang} in the dilute
regime of the Hubbard model in $d=1$.  In the $t-J$ model\cite{yang2} ($t=J$),
$\alpha_c$ starts at $2$ at $n=1$ and approaches a value of $4$ at
$n=0$.  Note, $\alpha_c=4$ is the non-interacting value.   That is, in
$d=1$ in the dilute regime,
density correlations decay as $r^{-2}$ in the $U\rightarrow\infty$ Hubbard
model and as $r^{-4}$ in the $t-J$ (t=J).  This discrepancy is a clear indicator that relevant
low-energy physics is lost if double occupancy of bare electrons is projected out in the parameter range considered here.
Supposedly, this is captured by the $t/U$ corrections to the electron operators
in Eq. (\ref{trans}).  In the procedure we outline in the Section III, we show
that all of the physics described by the string of operators in the
$t/U$ corrections (Eq. (\ref{trans}) is described by a single charge 2e bosonic field.

\begin{table}
\caption{\label{tjsumm}Comparison between Hubbard and t-J models.
The exponents $\alpha_c$ and $\theta$ were computed using Bethe ansatz for
the supersymmetric ($t=J$) $t-J$ and Hubbard models in $d=1$. Respectively, these exponents govern the asymptotic form of the density correlations and the momentum distribution functions.
$Z_{\rm 1h}$ denotes the quasiparticle weight for a single hole in
Mott insulator described by either the Hubbard or t-J models,
$\sigma(T=0,n>0.9)$ the conductivity for fillings exceeding $n=0.9$,
and Mid-IR denotes the mid-infrared band in the optical conductivity}
\footnotesize\rm
\begin{tabular*}{\textwidth}{@{}l*{15}{@{\extracolsep{0pt plus12pt}}l}}
\br                                
& ${\rm t-J}$ &${\rm Hubbard} $\\
\mr
$\alpha_c(n=0)$ & 4\cite{yang2}& 2\cite{yang}\\
$\theta$& $(\alpha_c-4)^2/16\alpha_c$\cite{yang2}&1/8 \quad $U\rightarrow\infty$\cite{yang}\\
$Z_{\rm 1 h}$ & ${\rm finite}$\cite{shole1,shole2,shole22} &
$L^{-\alpha}\cite{shole3}$\\
{\rm Mid-IR} & {\rm none}\cite{haulek,zaanen} & {\rm
yes}\cite{millis,chakrab}\\
{\rm Luttinger surface at n=1}&${\rm none}$\cite{dzy,zeros,essler} &
{\rm yes}\cite{zeros,konstanzeros,essler}\\
\br
\end{tabular*}
\end{table}

\section{New Theory: Hidden Charge 2e Boson}

The Wilsonian program for constructing a low-energy theory is to
explicitly integrate over the fields at high energy.  The theory that
results from this procedure should contain all the physics at low
energy.  In the context of the Hubbard model, it should explicitly
tell us that $L/n_h>1$, a key defining feature of a gapped phase
without symmetry breaking.  We now show how this comes about.

For the
Hubbard model, one has to cleanly associate the
physics on the energy scale $U$ with an elemental field that can be
integrated out either by using fermionic or bosonic path-integral
methods. The Hubbard model in its traditional form does not admit such
a treatment.  However, one can bring the Hubbard model into the
appropriate form by introducing an elemental field that describes the
excitations far away from the chemical potential.  For hole-doping,
this constitutes the excitations in the upper-Hubbard band. In our
construction, we will extend the Hilbert space of the Hubbard model to
include an extra degree of freedom which will represent the upper
Hubbard band.  By a constraint, the new field will represent the
creation of double occupancy.  This field will enter the Lagrangian
with a mass of $U$.  When the constraint is solved, we recover
the Hubbard model.  However, since the new field has a mass of $U$,
the exact low-energy theory is obtained by integrating over this
field rather than by solving the constraint.  Consequently, the new
low-energy theory will contain an extra degree of freedom having to do
with the coupling with the high energy scale.    Since the constraint
field has to do with double occupancy, it must have charge 2e.

\subsection{Bohm/Pines Redux}
There is a great similarity between our treatment of the new
collective mode in the Hubbard model and the Bohm-Pines derivation of plasmons.
Here we briefly introduce the approach used by Bohm and Pines\cite{bohm} to describe the collective excitation
of the interacting electron gas. Shankar and Murthy\cite{sm} also used a
similar approach in their dipole analysis of the $\nu=1/2$ quantum
Hall state. The basic idea is to re-express the Hamiltonian in such a way that the
long-range part of the Coulomb interactions between the electron is described in terms of collective fields (plasma mode)
by enlarging the original Hilbert space. After we remove the unphysical states by a constraint, the resultant Hamiltonian
will transform to an interacting electron gas with only short-range Coulomb interactions coupled to a plasma oscillating mode.

The starting point is the general interacting electron Hamiltonian,
\beq
H&=&\sum_i \frac{p_i^2}{2m}+2\pi e^2 \sum_{kij} \frac{e^{i\vec k \cdot (\vec x_i -\vec x_j)}}{k^2}-2\pi n e^2 \sum_k \frac{1}{k^2}\nonumber
\eeq
where $n$ is the total electron density, the first term corresponds to the kinetic energy of the electrons, the second term is their Coulomb interaction and
the
third term is the self-energy which represents a uniform positive charge background.

The collective mode can be described by first enlarging the Hilbert space of the original electron gas to include a new
set of canonical variables, $(\pi_k, \theta_k)$ such that $[\theta_k, \pi_{k'}]=i\hbar \delta_{k,k'}$. The original Hamiltonian
becomes,
\beq
\label{H-plasma}
H&=&\sum_i \frac{p_i^2}{2m}-2\pi n e^2 \sum_k \left( \frac{1}{k^2}\right)\nonumber\\
&+&\frac{\sqrt{4\pi} e}{m}\sum_{ik} \vec \epsilon_k \cdot (\vec p_i -\hbar \vec k/2)\theta_k e^{i\vec k\cdot \vec x_i}
-\sum_k \frac{1}{2}\pi_k \pi_{-k} \nonumber\\
&+&\frac{2\pi e^2}{m}\sum_{ikl}
\vec\epsilon_k \cdot \vec\epsilon_l \theta_k \theta_l e^{i(\vec k+\vec l)\cdot \vec x_i}
\eeq
Here, $\vec\epsilon_k$ is the unit vector along the $\vec k$
direction. The relevant equation can be derived by rewriting the Hamiltonian as a non-interacting
electron system coupled to an external electric field, $\vec E(\vec x)$,
\beq
H=\sum_i \frac{1}{2m} \left( p_i - \frac{e}{c} A_i (\vec x)\right)^2 + \frac{1}{8\pi} \vec E(\vec x)^2,
\eeq
such that
\beq
\vec E(\vec x)=\sqrt{4\pi}\sum_k \pi_{-k} \vec \epsilon_k e^{i\vec k\cdot \vec x}.
\eeq
The corresponding longitudinal vector potential $\vec A(\vec x)$ is
\beq
\vec A(\vec x)=\sqrt{4\pi c^2} \sum_k \theta_k \vec \epsilon_k e^{i\vec k\cdot\vec x}.
\eeq
Here, both $\vec A(\vec x)$ and $\vec E(\vec x)$ are real and the unphysical states can be removed by the constraint,
\beq
\Omega_k = \pi_k -i \left ( \frac{4\pi e^2}{k^2}\right)^{\frac{1}{2}} \sum_i e^{-i\vec k\cdot \vec x_i}=0  \ \forall{\rm k},
\eeq
which was obtained by equating the electric field energy, $ \vec E(\vec x)^2/8\pi$ with the electron-electron interaction energy.
For the last term in Eq. (\ref{H-plasma}), the dominant part comes from $\vec k=-\vec l$. By defining the plasma frequency
\beq
\omega_p^2 =\frac{4\pi n e^2}{m},
\eeq
we are able to simplify the Hamiltonian
%%\begin{widetext}
\beq
H&=&\sum_i \frac{p_i^2}{2m}-2\pi n e^2 \sum_k \left( \frac{1}{k^2}\right)
+\frac{\sqrt{4\pi} e}{m}\sum_{ik} \vec \epsilon_k \cdot (\vec p_i -\hbar \vec k/2)\theta_k e^{i\vec k\cdot \vec x_i}
\nonumber\\
&-&\frac{1}{2}\sum_k \left( \pi_k\pi_{-k} +\omega_p^2 \theta_k 
\theta_{-k}\right),
\eeq
%%\end{widetext}
which describes the non-interacting electron gas coupled with the
plasma mode of frequency $\omega_p$. Here, we have simplified the derivation
by assuming the collective modes can oscillate with any frequency. In
a realistic system, a maximum cutoff frequency, $k_c$, determined by the
electron density, arises so that only the long-range electron-electron interaction can be transformed into the plasma mode, and
the electron gas retains a short-range Coulomb interaction. The magnitude of $k_c$ can be determined self-consistently by minimizing the
total energy. To summarize, we have mapped the original electron-electron interacting Hamiltonian to a non-interacting electron gas coupled to the plasma mode.
The key trick that made this happen was enlarging
the original Hilbert space.
\begin{figure}
\centering
\includegraphics[width=12.0cm]{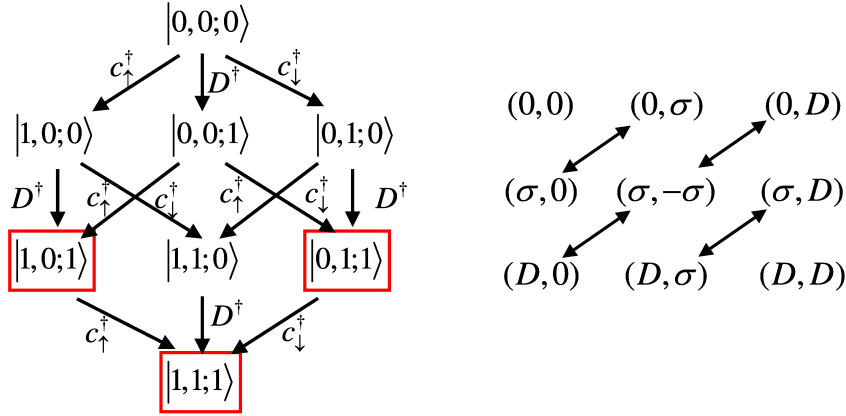}
\caption{a) Extended Hilbert space which allows a clean
 coarse-graining on the energy scale $U$. b) Hopping processes which
 are included in the Lagrangian in the extended Hilbert space.}
\label{exthops}
\end{figure}

\subsection{Charge 2e Boson}

The essence of the Bohm-Pines\cite{bohm} derivation is that plasmons,
as an independent degree of freedom, are only apparent when the
constraint is relaxed in the extended Hilbert space. As we will see,
the same is true for a doped Mott insulator. To this end\cite{ftm1,ftm2}, we  extend the Hilbert space $\otimes_i \left({\cal
F}_\uparrow\otimes {\cal F}_\downarrow\otimes {\cal F}_D\right)$,
where $\cal F$ denotes a fermionic Fock space.  In the left frame of
Fig. (\ref{exthops}), we indicate the states of the extended Hilbert
space for a single site.  The new extended states are shown in red.
Such states will be removed by a suitably chosen constraint which will associate $D_i^\dagger$ with the creation of
double-occupation.  In order to limit
the Hilbert space to single occupation in the $D$ sector, we will take
$D$ to be fermionic.  The field $D$ will enter the theory
as an elemental field with a large (order $U$) quadratic term and
precise interactions with the electronic degrees of freedom; the
low-energy (IR) theory is obtained by integrating out $D$. Because $D$
is fermionic, a trick is required to associate it with the creation of
double occupancy which clearly transforms as a boson. Essentially, we
will have to fermionize double occupancy.  This can be done by
imposing a constraint on the $D_i$ field such that 
\beq
D_i\approx \theta c_{i\uparrow}c_{i\downarrow}
\eeq
where $\theta$ is a Grassmann field which is normalized as
\beq
\int d^2\theta\ \bar{\theta}\theta=1.
\eeq  
The constraint will be imposed
through the use of a $\delta-$function and hence will enter the action
upon a subsequent exponentiation as in the implementation of the
constraint in the non-linear $\sigma-$model.  While
there is some similarity between $\theta$ and a
super-coordinate, this association is strictly formal in our case as
any dependence on the Grassmann parameter $\theta$ disappears.

For hole-doping, it is the upper Hubbard band that must be integrated
out.  The appropriate hopping processes that must be included in the
Lagrangian are given in the right-hand panel of Fig. (\ref{exthops}).
The Euclidean Lagrangian in the extended Hilbert space
which describes the hopping processes detailed above can be written
\beq\label{LE}
{\cal L}&&=\int d^2\theta\left[\bar{\theta}\theta\sum_{i,\sigma}(1- n_{i,-\sigma}) c^\dagger_{i,\sigma}\dot c_{i,\sigma} +\sum_i D_i^\dagger\dot D_i\right.\nonumber\\
&&+U\sum_j D^\dagger_jD_j- t\sum_{i,j,\sigma}g_{ij}\left[ C_{ij,\sigma}c^\dagger_{i,\sigma}c_{j,\sigma}
+D_i^\dagger c^\dagger_{j,\sigma}c_{i,\sigma}D_j\right.\nonumber\\
&&+\left.\left.(D_j^\dagger \theta c_{i,\sigma}V_\sigma c_{j,-\sigma}+h.c.)\right]+H_{\rm con}\right].
\eeq
Here, $g_{ij}$ selects out nearest neighbours (note that if we wanted to include next-to-nearest neighbour interactions, we need just modify the matrix $g_{ij}$ accordingly), the parameter $V_\sigma$
has values $V_\uparrow =1$, $V_\downarrow=-1$, and simply ensures that
$D$ couples to the spin singlet and the operator $C_{ij,\sigma}$ is of
the form
$C_{ij,\sigma}\equiv\bar\theta\theta\alpha_{ij,\sigma}\equiv\bar\theta\theta(1-n_{i,-\sigma})(1-n_{j,-\sigma})$ with number operators
$n_{i,\sigma}\equiv c^\dagger_{i,\sigma}c_{i,\sigma}$.
Note that the dynamical terms that appear in the Lagrangian exclude,
as they must, those sites already singly occupied, as they describe
the dynamics in the LHB.  The constraint Hamiltonian $H_{\rm con}$ is taken to be
\beq\label{con}
H_{\rm con} = s\bar{\theta}\sum_j\varphi_j^\dagger (D_j-\theta c_{j,\uparrow}c_{j,\downarrow})+h.c.,
\eeq
where $\varphi$ is a complex charge $2e$ bosonic field which enters the
theory as a Lagrange multiplier.  The constant $s$ has been inserted to
carry the units of energy.
It could be absorbed into the definition of the constraint field
$\varphi$.  
%However, this would change the commutation relations. 
There is a natural connection between $\varphi_i$ in our
theory and $\sigma$ in the non-linear $\sigma$-model.  Both start
their lives as Lagrange multipliers but both end up affecting the
low-energy physics.
At this point, there is some ambiguity in the normalization of
$\varphi$, but we expect that this will be set dynamically. We will find
that if a true infrared limit exists, then $s$ must be of order the
hopping matrix element $t$.  

Now, as remarked previously, we have chosen the Lagrangian (\ref{LE}) so
that this theory is equivalent to the Hubbard model. To demonstrate
this, we first show that once the constraint is solved, we obtain the
Hubbard model. Hence, the Lagrangian we have formulated is the Hubbard
model written in a non-traditional form -- in a sense, we have
inserted unity into the Hubbard model path integral in a rather
complicated fashion.  To this end, we compute the partition function
\beq\label{Z}
Z=\int [{\cal D}c\ {\cal D}c^\dagger\ {\cal D}D\ {\cal D}D^\dagger\ 
{\cal D}\varphi\ {\cal D}\varphi^\dagger]\exp^{-\int_0^\tau{\cal L} dt}.
\eeq
with ${\cal L}$ given by (\ref{LE}). We note that $\varphi$ is a Lagrange
multiplier.   As shown previously\cite{ftm2}, in the Euclidean
signature, the fluctuations of the real and imaginary parts of
$\varphi_i$ must be integrated along the imaginary axis for $\varphi_i$
to be regarded as a Lagrangian multiplier.   The $\varphi$ integrations
(over the real and imaginary parts) are precisely a representation of (a
series of) $\delta$-functions of the form,
\beq
\delta\left(\int d\theta D_i-\int d\theta\ \theta c_{i,\uparrow}c_{i,\downarrow}\right).
\eeq

If we wish to recover the Hubbard model, we need only to integrate over
$D_i$, which is straightforward because of the $\delta$-functions.  The
dynamical terms yield
\beq
&&\int d^2\theta\ \bar\theta\theta\left[\sum_{i,\sigma}(1- n_{i,-\sigma}) 
c^\dagger_{i,\sigma}\dot c_{i,\sigma}+\sum_i c^\dagger_{i,\downarrow}
c_{i,\uparrow}^\dagger\dot c_{i,\uparrow}c_{i,\downarrow}\right.\nonumber\\
&&\qquad\qquad\left.+\sum_ic^\dagger_{i,\downarrow}c_{i,\uparrow}^\dagger 
c_{i,\uparrow}\dot c_{i,\downarrow}\right]\nonumber\\
&&=\int d^2\theta\ \bar\theta\theta\sum_{i,\sigma}\left[(1-n_{i,-\sigma})
c^\dagger_{i,\sigma}\dot c_{i,\sigma}+n_{i,-\sigma}c_{i,\sigma}^\dagger
\dot c_{i,\sigma}\right]\nonumber\\
&&=\int d^2\theta\ \bar\theta\theta\sum_{i,\sigma}c_{i,\sigma}^\dagger \dot c_{i,\sigma}.
\eeq
Likewise the term proportional to $V_\sigma$ yields
\beq\label{den1}
&&\int d^2\theta\ \bar\theta\theta \sum_{i,j}g_{ij}\left[c_{j,\downarrow}^\dagger 
c_{j,\uparrow}^\dagger(c_{i,\uparrow}c_{j,\downarrow}-c_{i,\downarrow}c_{j,\uparrow})\right]
+h.c.\nonumber\\
&&=\int d^2\theta\ \bar\theta\theta \sum_{i,j,\sigma}g_{ij}n_{j,-\sigma}
c^\dagger_{j,\sigma}c_{i,\sigma}+h.c.
\eeq
Finally, the two $D$ field terms give rise to
\beq\label{den2}
&&\int d^2\theta\ \bar\theta\theta\sum_{i,j}g_{ij}\left[c_{i,\downarrow}^\dagger c_{i,\uparrow}^\dagger(c_{j,\uparrow}^\dagger c_{i,\uparrow}+c_{j,\downarrow}^\dagger c_{i,\downarrow})c_{j,\uparrow}c_{j,\downarrow}\right]\nonumber\\
&&=-\int d^2\theta\ \bar\theta\theta\sum_{i,j}g_{ij}n_{j,-\sigma}n_{i,-\sigma}c_{i,\sigma}^\dagger c_{j,\sigma}.
\eeq
Eqs. (\ref{den1}) and (\ref{den2}) add to the constrained hopping term
in the Lagrangian (the term proportional to $C_{ij,\sigma}$) to yield the
standard kinetic energy term in the Hubbard model. Finally, the
$D^\dagger D$ term  generates the on-site repulsion of the Hubbard
model.  Consequently, by integrating over $\varphi_i$ followed by an
integration over $D_i$, we recover the Lagrangian,
\beq
\int d^2\theta\ \bar\theta\theta {\cal L}_{\rm Hubb}=\sum_{i,\sigma}c_{i,\sigma}^\dagger\dot c_{i,\sigma}+H_{\rm Hubb},
\eeq
of the Hubbard model.  This constitutes the ultra-violet (UV) limit of
our theory. In this limit, it is clear that the Grassmann variables
amount to an insertion of unity and hence play no role.  Further, in
this limit the extended Hilbert space contracts, unphysical states such
as  $|1,0,1\rangle$, $|0,1,1\rangle$, $|1,1,1\rangle$ are set to zero,
and we identify $|1,1,0\rangle$ with $|0,0,1\rangle$.  Note there is no
contradiction between treating $D$ as fermionic and the constraint in
Eq. (\ref{con}). The constraint never governs the commutation relation
for $D$. The value of $D$ is determined by Eq. (\ref{con}) only when
$\varphi$ is integrated over. This is followed immediately by an
integration over $D$ at which point $D$ is eliminated from the theory.

The advantage of our starting Lagrangian over the traditional writing of
the Hubbard model is that we are able to coarse grain the system cleanly
for $U\gg t$.  The energy scale associated with $D$ is the large on-site
energy $U$.  Hence, it makes sense, instead of solving the constraint,
to integrate out $D$.  The resultant theory will contain explicitly the
bosonic field, $\varphi$.   As  a result of this field, double occupancy
will remain, though the energy cost will be shifted from the UV to the
infrared (IR).  Because the theory is Gaussian, the integration over
$D_i$ can be done exactly.  This is the ultimate utility of the
expansion of the Hilbert space -- we have isolated the high energy
physics into this Gaussian field. As a result of the dynamical term in
the action, integration over $D$ will yield a theory that is frequency
dependent. The frequency will enter in the combination $\omega+U$ which
will appear in  denominators.  Since $U$ is the largest energy scale, we
expand in powers of $\omega/U$; the leading term yields the proper
$\omega=0$ low-energy theory.   Since the theory is Gaussian, it
suffices to complete the square in the $D$-field. To accomplish this, we
define the matrix
\beq\label{eom}
{\cal M}_{ij}=\delta_{ij}-\frac{t}{(\omega+U)}g_{ij}
\sum_\sigma c_{j,\sigma}^\dagger c_{i,\sigma}
\eeq
and $b_{i}=\sum_{j}b_{ij}=\sum_{j,\sigma} g_{ij}c_{j,\sigma}V_\sigma
c_{i,-\sigma}$. At zero frequency the Hamiltonian is
\beq
H^{IR}_h = -t\sum_{i,j,\sigma}g_{ij} \alpha_{ij,\sigma}
c^\dagger_{i,\sigma}c_{j,\sigma}+ H_{\rm int}-\frac{1}{\beta}Tr\ln{\cal M},
\nonumber
\eeq
where
%\beq\label{HIR}
%H_{\rm int}=-\frac{t^2}U \sum_{j,k} b^\dagger_{j} ({\cal M}^{-1})_{jk} b_{k}-\frac{s^2}U\sum_{i,j}\varphi_i^\dagger
% ({\cal M}^{-1})_{ij} \varphi_j\nonumber\\
%-s\sum_j\varphi_j^\dagger c_{j,\uparrow}c_{j,\downarrow}-\frac{st}U \sum_{i,j}\varphi^\dagger_i ({\cal M}^{-1})_{ij}
%b_{j}+h.c.\;\;\nonumber\\
%-\frac{1}{\beta}Tr\ln{\cal M}^{-1},
%\eeq
\beq\label{HIR}
H_{\rm int}&=&-\frac{t^2}U \sum_{j,k} b^\dagger_{j} ({\cal M}^{-1})_{jk} b_{k}
-\frac{s^2}U\sum_{i,j}\varphi_i^\dagger
({\cal M}^{-1})_{ij} \varphi_j\nonumber\\
&-&s\sum_j\varphi_j^\dagger c_{j,\uparrow}c_{j,\downarrow}
+\frac{st}U \sum_{i,j}\varphi^\dagger_i ({\cal M}^{-1})_{ij}
b_{j}+h.c.
\eeq
which constitutes the true (IR) limit as long as the energy scale $s$ is
not of order $U$.  As we have retained all powers of $t/U$,
Eq. (\ref{HIR}) is exact. If $s\sim O(U)$ then we should also integrate out
$\varphi_i$ -- this integration is again Gaussian and can be done
exactly; one can easily check that this leads precisely back to the UV
theory, the Hubbard model.  Hence, the only way in which a
low-energy theory of the Hubbard model exists is if the energy scale for
the dynamics that $\varphi$ mediates is $O(t)$.   This observation is
significant because it lays plain the principal condition for the
existence of an IR limit of the Hubbard model.  Since the order of
integrations we have performed here does not matter, integration over
$\varphi_i$ in the path integral for the action corresponding to Eq. 
(\ref{HIR}) also yields the Hubbard model.  As we have shown
elsewhere\cite{ftm2} the Wick rotation must be taken into
consideration to complete the Gaussian integration over $\varphi_i$.
Finally, as we have shown previously\cite{ftm2}, the theory derived
here could easily have been constructed in terms of the Hubbard
operators, $\xi$ and $\eta$.  This offers no further complication.  In
so doing, the spin-spin interaction which arises from the first term
in Eq. (\ref{HIR}) would have been projected onto a subspace which
prohibits double occupancy.  That is, it would be equivalent to the
spin-spin interaction in the standard $t-J$ model.  Since double occupancy still survives at
low energies and is mediated by the $\varphi_i$ terms, such a
rewriting of the low-energy Hamiltonian is strictly optional.

To fix the constant $s$, we determine how the electron operator transforms in
the exact theory. As is standard, we add a source term to the
starting Lagrangian which generates the canonical electron operator
when the constraint is solved. For hole-doping, the appropriate
transformation that yields the canonical electron operator in the UV
is
\beq
{\cal L}\rightarrow {\cal L}+\sum_{i,\sigma} J_{i,\sigma}\left[\bar\theta\theta(1-n_{i,-\sigma} ) c_{i,\sigma}^\dagger + V_\sigma D_i^\dagger \theta c_{i,-\sigma}\right] +
h.c.\nonumber
\eeq
However, in the IR in which we only integrate over the heavy degree of
freedom, $D_i$, the electron creation operator becomes
\beq\label{cop}
c^\dagger_{i,\sigma}&\rightarrow&(1-n_{i,-\sigma})c_{i,\sigma}^\dagger
+ V_\sigma \frac{t}{U} b_i c_{i,-\sigma}
+ V_\sigma \frac{s}{U}\varphi_i^\dagger c_{i,-\sigma}
\eeq
to linear order in $t/U$. This equation resembles the
transformed electron operator in Eq. (\ref{trans}).   In
fact, the first two terms are identical. The last term in
Eq. (\ref{trans}) is associated with double occupation. In
Eq. (\ref{cop}), this role is played by $\varphi_i$. Demanding that Eqs. (\ref{trans}) and (\ref{cop}) agree requires that $s= t$, thereby eliminating
any ambiguity associated with the constraint
field. Consequently, the complicated interactions appearing in
Eq. (\ref{trans}) as a result of the inequivalence between
$c_{i\sigma}$ and $f_{i\sigma}$ are replaced by a single charge $2e$ bosonic field
$\varphi_i$ which generates dynamical spectral weight transfer across the
Mott gap. The interaction in Fig. (\ref{dos}), corresponding to the
second-order process in the term $\varphi_i^\dagger b_i$, is the key
physical process that enters the dynamics at low-energy.  That the dynamical spectral weight transfer can be captured
by a charge $2e$ bosonic degree of freedom is the key outcome of the
exact integration of the high-energy scale. This bosonic field represents a
collective excitation of the upper and lower Hubbard bands.  Hence, we
have successfully unearthed the extra degree of freedom associated
with $L/n_h>1$ physics in a doped Mott insulator.  

\subsection{Half-Filling: Mott gap and antiferromagnetism}

There is an important simplification that obtains at half-filling that
points to one of the potential uses of this theory: the dynamical mode that generates the Mott gap.  Recall our ultimate
task was to integrate out the degrees of freedom far away from the
chemical potential.  At half-filling, both the lower and upper Hubbard
bands are at high energy.  Hence, both must be integrated out to
obtain the true low-energy theory.  However, at present, we have only
integrated out the UHB.  Because the integration of the UHB is not
equivalent to simply integrating out double occupancy, a trivial
particle-hole transformation does not help us to determine how the
low-energy theory should be formulated in this case. 

Ultimately we  have to introduce two new fermionic fields $D$ and $\tilde D$
associated with the double occupancy and double holes, respectively.
To proceed, we consider the Lagrangian,
\beq\label{hfuv}
{\cal L}^{\rm hf}_{\rm UV}& =&\int d^2\theta\left[ iD^\dagger\dot D+i\tilde
  D\dot{\tilde D}^\dagger-\frac{U}{2}(D^\dagger D-\tilde D\tilde
  D^\dagger)\right.\nonumber\\
&+&\left. \frac{t}{2}D^\dagger\theta b +\frac{t}{2}\bar\theta b\tilde
  D+h.c.+ s\bar\theta\varphi^\dagger (D-\theta c_\uparrow
  c_\downarrow)\right.\nonumber\\
&+&\left. \tilde s\bar\theta\tilde\varphi^\dagger (\tilde D-\theta c^\dagger_\uparrow c^\dagger_\downarrow)+h.c.\right],
\eeq
contains the two constraint charge $\pm 2e$ bosonic fields, $\varphi^\dagger_i$
(charge $2e$)
and $\tilde\varphi^\dagger_i$ (charge $-2e$) which enter the theory as
Lagrange multipliers for the creation of double occupancy and double
holes, respectively. Similar to the previous
result, if we first integrate out both the bosonic fields
$\varphi_i$, $\tilde \varphi_i$ and then $D_i$, $\tilde D_i$, the
Hubbard model is obtained and the generalised theory Eq.\ref{hfuv} yields the correct UV limit. In deriving the UV limit, it is crucial that the operators representing the creation of double occupancy and double holes remain in the order indicated in the Lagrangian.  However, a different
IR limit is obtained if we first integrate out $D_i$ and $\tilde
D_i$,
\beq\label{hfir}
{\cal L}^{\rm hf}_{\rm IR}&=&
-\left(s\varphi^\dagger+\frac12 t b^\dagger\right)L_-^{-1}\left(s^*\varphi+\frac12 t b\right)\nonumber\\
&+&\left(\tilde s^*\tilde\varphi+\frac12 t b^\dagger\right)L_+^{-1}\left(\tilde s\tilde\varphi^\dagger+\frac 12 tb\right)\nonumber\\
&-&\left(s\varphi^\dagger-\tilde s^*\tilde \varphi\right) c_\uparrow c_\downarrow+h.c.,
\eeq
where  $L_\pm=i\frac{d}{d t}\pm \frac{U}{2}$.  This theory is exact
and hence should contain the source of the Mott gap. 
This Lagrangian is invariant under the transformation
$c_{i,\sigma}\rightarrow \exp(i \vec Q \cdot \vec R_i)
c_{i,\sigma}^\dagger$, $\varphi_i \leftrightarrow \tilde \varphi_i$ and
$s \leftrightarrow \tilde s$. This invariance reflects the symmetry
between the double occupancy and the double hole in the system at half
filling. In contrast to the doped case as in Eq. (\ref{HIR}), no $\cal M$ matrices appear in the IR theory at half
filling.  Consequently, we arrive at a closed form for the low-energy
theory at half-filling in which no bare field has dynamics. The $b^\dagger b + b b^\dagger$ terms include a
spin-spin interaction as well as a three-site hopping term. However, at
half-filling, the three-site hopping term vanishes. 

It is interesting to note that Eq. (\ref{hfir}), as an exact
low-energy theory of the Hubbard model, is not equivalent to the
Heisenberg model.   Only the $b^\dagger b$ terms yield the Heisenberg
model. Hence, we consider the separation,
\beq
{\cal L}_{\rm Mott}={\cal L}_{\rm hf}^{\rm IR}+\frac14 t^2
b^\dagger L_-^{-1}b+\frac14 t^2 b^\dagger L_+^{-1}b
\eeq
which explicitly removes the spin-spin term from the low-energy
Lagrangian.  As will be seen, the dynamics leading to the Mott gap can
be constructed entirely from ${\cal L}_{\rm Mott}$.   That ${\cal
  L}_{\rm hf} ^{\rm IR}$ is not equivalent to the Heisenberg model is
not surprising for three related reasons.  First, explicit evaluation of the path integral over the new degrees of freedom must regenerate
the original Hubbard model. Hence, there must be something left over
once we subtract the Heisenberg terms from the low-energy theory of
the Hubbard model. Second, as pointed out previously\cite{lsu21,lsu22}, the
Heisenberg model has a local SU(2) symmetry not the global SU(2)
symmetry of the Hubbard model.  Hence, the true low-energy theory of
the Hubbard model at half-filling must have other terms that break the
local SU(2) and reinstate 
the global SU(2) symmetry.  As we have shown previously\cite{ftm2}, all of the
terms involving the $\varphi_i$ and $\tilde\varphi_i$ degrees of
freedom restore the global SU(2) symmetry. Consequently, the emergence
of the new local SU(2) symmetry is a function entirely of
projection onto the singly-occupied electron subspace.  Third, at half-filling, the Hubbard model
possesses a surface of zeros\cite{dzy,zeros} of the single-particle Green function along a connected surface in momentum space,
whereas the Heisenberg model does not. The absence of such a surface
of zeros, the Luttinger surface, is also a function of projection.  In
fact, all of these three failures of the Heisenberg model arise from
hard projection, which Eq. (\ref{trans}) shows is not correct even at
half-filling.  The
non-trivial implication of the zero surface is that the real part of the
Green function,
\beq\label{green}
R_\sigma(\vec k,0)=-\int_{-\infty}^{-\Delta_-}d\epsilon'\frac{ A_\sigma(\vec
k,\epsilon')}{\epsilon'}-\int_{\Delta_+}^\infty d\epsilon'
\frac{A_\sigma(\vec k,\epsilon')}{\epsilon'}
\eeq
vanishes. Here $A_\sigma(k,\epsilon)$ is the single-particle
spectral function which we are assuming to have a gap of width
$2\Delta$ symmetrically located about the chemical potential at
$\epsilon=0$.  Because $A(\vec k,\epsilon)>0$ away from the gap, and
$\epsilon$ changes sign above and below the gap, Eq. (\ref{green})
can pass through zero.  For this state of affairs to obtain, the
pieces of the integral below and above the gap must be retained.
Projected models which throw away the UHB fail to recover the zero
surface. 

\subsubsection{Mott Gap}

What Eq. (\ref{hfir}) makes clear is that all the
information regarding the surface of zeros is now encoded into the
bosonic fields $\varphi_i$ and $\bar\varphi_i$. 
While numerical methods\cite{cluster1,cluster2} exist which lead
to a Mott gap, there has been no explicit
demonstration of the dynamical degrees of freedom that ultimately
generate this gap.  The bosonic degrees of freedom in
Eq. (\ref{hfir}) solve this problem.

To proceed, we  transform to frequency and momentum space and focus on
a square lattice as in the cuprates.  Defining $\varphi(t) = \int
d\omega\ e^{-i\omega t}\varphi_\omega$, the energy dispersion,
$\varepsilon^{(\vec k)}_{\vec p}=4\sum_\mu\cos(k_\mu a/2)\cos(p_\mu
a)$, and the Fourier
transform of $b_{i}$,
\beq\label{fb}
b_{\vec k}=\sum_{\vec p}\varepsilon^{(\vec k)}_{\vec p}\ c_{\vec k/2+\vec p,\uparrow}c_{\vec k/2-\vec p,\downarrow},
\eeq
we arrive at the exact working expression,
\beq\label{lfreq}
{\cal L}^{\rm hf}_{\rm IR}&=&-\frac{|s|^2}{(\omega-U/2)}\varphi_{\omega,\vec
  k}^\dagger\varphi_{\omega,\vec k}
+\frac{|s|^2}{(\omega+U/2)}\tilde\varphi_{-\omega,\vec
  k}^\dagger\tilde\varphi_{-\omega,\vec k}\nonumber\\
&+& \frac{Ut^2}{U^2-4\omega^2} b_{\omega,\vec k}^\dagger b_{\omega,\vec k}\nonumber\\
&+&(s\alpha_{\vec p}^{(\vec k)}(\omega)\varphi^\dagger_{\omega,\vec k}+\tilde s^*\tilde\alpha_{\vec p}^{(\vec k)}(\omega)\tilde\varphi_{-\omega,\vec k})\nonumber\\&\times&(c_{\vec k/2+\vec p,\uparrow}c_{\vec k/2-\vec p,\downarrow})_\omega +h.c.
\label{eq:kinterms}
\eeq
for the low-energy Lagrangian where we have suppressed the implied
integration over frequency and introduced the coupling constants,
\beq
\alpha_{\vec p}^{(\vec k)}(\omega)=\frac{U-t\varepsilon_{\vec p}^{(\vec k)}-2\omega}{U-2\omega}\nonumber\\
\tilde\alpha_{\vec p}^{(\vec k)}(\omega)=\frac{U+t\varepsilon_{\vec
    p}^{(\vec k)}+2\omega}{U+2\omega}
\eeq
which play a central role in this theory.  They, in fact, will determine
the spectral weight in the lower and upper Hubbard bands,
respectively. 

Although Eq. (\ref{lfreq})  lacks any kinetic terms, an analysis
similar to that used by Polchinski\cite{polchinski} in the context of
Fermi liquids can be used. 
  The key point in that argument is that all renormalizations are
  towards the surface in momentum space where the spectral weight
  resides. In a theory of weakly coupled constituents, setting the
  coefficient of the quadratic terms in a Lagrangian would
  determine their dispersion. As is evident from Eq. (\ref{lfreq}),
  the coupling constants for all of the quadratic terms never
  vanish.  All else being equal, this implies that there is no spectral
  weight anywhere.  What we have shown\cite{unpub} is that the terms
  in which 
the bosonic
  degrees of freedom and the fermions are coupled determine where
  spectral weight resides. In fact, it is these terms that should
  properly be regarded as quadratic.  To this end, we make the transformation,
$\varphi_\omega\to \sqrt{1-2\omega/U}\ \varphi_\omega$, 
$\tilde\varphi_\omega\to \sqrt{1+2\omega/U}\ \tilde\varphi_\omega$, and
$(cc)_\omega\to \sqrt{1-4\omega^2/U^2}\ (cc)_\omega$
which recasts the Lagrangian as 
\beq
L^{\rm hf}_{\rm IR}&\to& 2\frac{|s|^2}{U}|\varphi_\omega|^2+2\frac{|\tilde s|^2}{U}|\tilde\varphi_{-\omega}|^2+ \frac{t^2}{U} |b_\omega|^2\label{eq:actlineone}\\
&&+s\gamma_{\vec p}^{(\vec k)}(\omega)\varphi^\dagger_{\omega,\vec k}c_{\vec k/2+\vec p,\omega/2+\omega',\uparrow}c_{\vec k/2-\vec p,\omega/2-\omega',\downarrow}\label{eq:actlineone}\\
&+&\tilde s^*\tilde\gamma_{\vec p}^{(\vec k)}(\omega)\tilde\varphi_{-\omega,\vec k}c_{\vec k/2+\vec p,\omega/2+\omega',\uparrow}c_{\vec k/2-\vec p,\omega/2-\omega',\downarrow}\nonumber\\&+&h.c.\label{eq:actlinetwo}
\eeq
where
\beq
\gamma_{\vec p}^{(\vec k)}(\omega)&=&\frac{-U+t\varepsilon_{\vec
    p}^{(\vec k)}
+2\omega}{U}\sqrt{1+2\omega/U}\nonumber\\
\tilde\gamma_{\vec p}^{(\vec k)}(\omega)&=&\frac{U+t\varepsilon_{\vec p}^{(\vec k)}+2\omega}{U}\sqrt{1-2\omega/U}.
\eeq

The vanishing of the coefficients $\gamma^{\vec k}_{\vec p}$ and $\tilde\gamma^{\vec
  k}_{\vec p}$ determine where the spectral weight lies. Consider initially $\vec k=0$ so that the
dispersion simplifies to $4\sum_\mu \cos p_\mu$.  When
$\omega=U/2$ $(\omega=-U/2)$, $\sum_\mu \cos p_\mu=0$ defines the
momentum surface
along which $\gamma_{\vec p}$ ($\tilde\gamma_\vec{p}$) vanishes.  This
surface corresponds to the diamond  $a\vec p=(ap_x,\pm\pi-ap_x)$ as
depicted in Fig. (\ref{mottgap}).  These features define the center
of the LHB ($-U/2$) and UHB ($U/2$).  At each momentum in the first
Brillouin zone (FBZ), spectral
weight develops at two distinct energies.  This state of affairs
obtains because $\gamma_{\vec p}=0$ between $U/2-4t\le\omega\le
U/2+4t$ and $\tilde\gamma_{\vec p}=0$ for $-U/2-4t\le\omega\le
-U/2+4t$ for each momentum state in the FBZ.   Within each energy range, the
associated operator, which is of the form $\varphi^\dagger c c$ (UHB) or
$\tilde\varphi cc$ (LHB), can
be viewed as a quadratic kinetic term.  The $(0,0)$ point
corresponds to the lowest energy state in each LHB and UHB, that is,
$\omega=-U/2-4t$ and $\omega=U/2-4t$ whereas $(\pi,\pi)$ sits at the
top of each band at $\omega=-U/2+4t$ and $\omega=U/2+4t$.  Consequently,
at each momentum, the splitting between the turn-on of the 
spectral weight in the UHB  and LHB is $U$.  When $U>W$, each
momentum state lacks spectral weight over a common range of energies.
As a consequence, a hard gap opens in the
spectrum.  This is
the Mott gap (for the composite excitations not the electrons), and
its origin is the emergence of composite excitations
described by the operators $\varphi^\dagger cc$ (UHB) and
$\tilde\varphi cc$ (LHB), which we loosely interpret as bound states. 
As our analysis thus far is exact, we
conclude that in the absence of any symmetry breaking,
the coefficients $\gamma_{\vec p}$ and
$\tilde \gamma_{\vec p}$ determine the dispersion for the excitations
  that comprise the here-to-fore undefined\cite{laughlin} UHB and LHB. Inclusion of
  the center-of-mass momentum $\vec k$ simply shifts the value of the
  momentum at which the dispersion changes sign, thereby keeping the
  Mott gap in tact. 
 \begin{figure}
\centering
\includegraphics[width=13.0cm,angle=0]{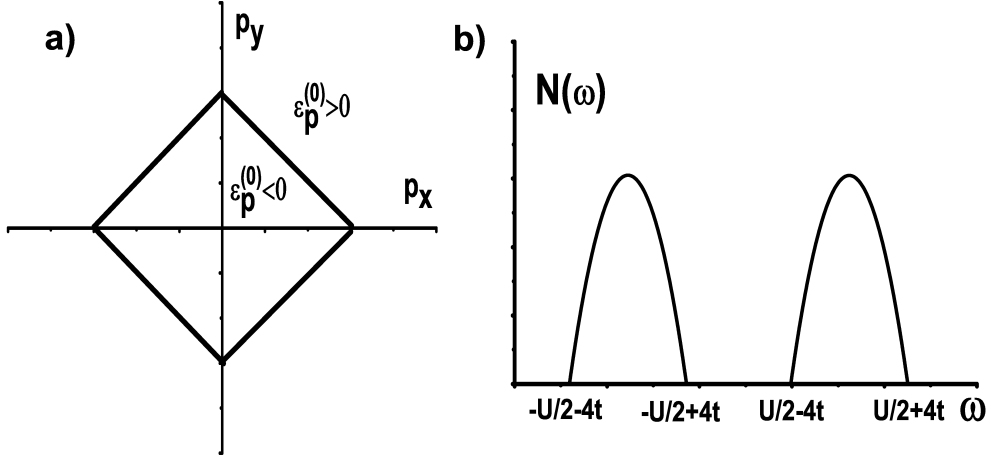}
\caption{a) Diamond-shaped surface in momentum space where the
  particle dispersion changes sign.  b) Turn-on of the spectral weight in the upper and lower Hubbard
  bands as a function of energy and momentum.   In the UHB, the
  spectral density is determined to $\gamma_{\vec p}$  while for the LHB it is
  governed by $\tilde\gamma_{\vec p}$.  The corresponding operators
which describe the turn-on of the spectral weight are the composite
excitations $\varphi^\dagger cc$ (UHB) and $\tilde\varphi cc$ (LHB). }  
\label{mottgap}
\end{figure}

 Thus far, we have established the Mott gap in terms of a set of
 composite excitations which are orthogonal in that they never lead to
 a turn-on of the spectral weight in the same range of energies.
Ultimately, we would like to know the
 spectral function in terms of the original electron degrees of
 freedom. The lack of any derivative terms 
in the action
with respect to $\varphi_i$ implies that we can treat $\varphi$ as a
spatially homogeneous field.  {\it A priori},
gradient terms with respect to $\varphi_i$ are possible.  However,
such terms are absent from the exact low-energy theory as such terms
would indicate the presence of a freely propagating bosonic degree of
freedom at half-filling.  It is precisely because such terms are
absent that we were able to identify that the only propagating degrees
of freedom at half-filling are gapped excitations. 

To proceed, we rewrite the coefficient of the boson-fermi terms as
\beq
\Delta(k,\omega,\phi,\tilde{\phi})&=&-s(\phi^{\dagger}-\tilde{\phi})\nonumber\\
&+&(\frac{st}{U-2\omega-i\delta}\phi^{\dagger}+\frac{st}{U+2\omega+i\delta}\tilde{\phi})\alpha(k)\nonumber
\eeq
and
\beq
\alpha(k)=2t(\cos(k_{x})+\cos(k_{y})).
\eeq
In this treatment, any non-trivial dynamics leading to the Mott gap will arise only from the the second term in $\Delta(k,\omega,\phi,\tilde\phi)$.  Performing the Wick rotation, $\phi \rightarrow  i\phi$ and 
$\phi^{*} \rightarrow i\phi^{*}$, we recast the single-particle electron Green
function as
\beq
G(k,\omega)=\int d\phi\int
d\tilde{\phi}G(k,\omega,\phi,\tilde{\phi})\exp^{-\int d\omega
{\cal  L}_{\rm Mott }}
\eeq
where
\beq
G(k,\omega,\phi,\tilde{\phi})=\frac{i\delta}{|\Delta(k,\omega,\phi,\tilde{\phi})|^{2}+i\delta}.
\eeq
At first glance, the Green function seems to have a vanishing imaginary
part. However, because of the $i\delta$ in the gap function, $\Delta(k,\omega,\phi,\tilde{\phi})$,
the imaginary part of the Green function
\beq
\Im G(k,\omega,\phi,\tilde{\phi})&=&\lim_{\delta\rightarrow 0}\left[(U-2\omega)^{2}+\delta^{2}\right]\left[(U+2\omega)^{2}+\delta^{2}\right]\nonumber\\
&\times&\frac{\delta}{A^{2}+\left(2A(\phi+\tilde{\phi})+B^{2}\right)\delta^{2}+O(\delta^{4})}\nonumber\\
&=&\frac{(U-2\omega)^{2}(U+2\omega)^{2}}{B}\delta(A)
\eeq
is explicitly non-zero.  Here 
\beq
A & = &
\left[U^{2}-4\omega^{2}-2\alpha_{k}(U+2\omega)\right]\phi+\left[U^{2}-4\omega^{2}-2\alpha_{k}(U-2\omega)\right]\tilde{\phi}\nonumber\\
B & = & 2\phi(2\omega+\alpha_{k})+2\tilde{\phi}(2\omega+\alpha_{k})
\eeq
To complete the calculation, we performed the $\varphi_i$ and
$\tilde\varphi$ integrations numerically.  The results in
Fig. (\ref{mottgap}) clearly show that a Mott gap exists and the
spectral weight is momentum dependent.  At $(0,0)$, the
spectral weight lies predominantly in the LHB whereas at $(\pi,\pi)$
it lies in the UHB.  Consequently, the real part of the Green function
must change sign on some momentum surface between these two limits.
The location of the zero surface is the Fermi surface of the
non-interacting system as it must be for the half-filled system with
particle-hole symmetry.
We find then that the Mott gap arises from the dynamics of the two charge
$|2e|$ bosonic fields.  This is the first time the Mott gap has been
derived dynamically, in particular by a collective degree of freedom
of the lower and upper Hubbard bands. Relative to the gap in the
spectrum for the composite excitations that diagonalise the 
   fermion-boson terms in Eq. (\ref{eq:actlinetwo}), the gap in the electron
spectrum is larger.  This is not surprising as the bare electrons do
   not have unit overlap with the composite excitations.   In
   addition, the momentum dependence of the
spectral function is identical to that obtained by dynamical mean-field
  calculations\cite{cluster1} thereby lending crecedence to
   such cluster\cite{mg1}.  
\begin{figure}
\centering
\includegraphics[height=8.0cm,angle=90]{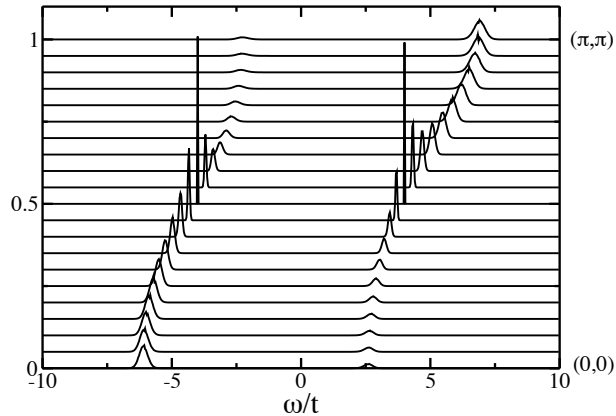}
\caption{Electron spectral function corresponding to ${\cal L}_{\rm
    Mott}$ for $U=8t$.
 The gap here is generated entirely from the dynamics of the charge
 2e bosonic fields that emerge from integrating out the upper and
 lower Hubbard bands at half-filling.}
\label{mottgap}
\end{figure}

An open question that this analysis provokes is whether or not the
turn-on of the spectral weight in a Mott insulator is governed by a fixed point.  If
so, then in analogy with the Fermi liquid analysis\cite{polchinski},
all the interactions except those that govern the turn-on of the spectral weight
should be irrelevant.  That is, the $\varphi^\dagger cc$ and the
$\tilde\varphi cc$ terms represent a natural theory.  Indeed, the
analysis presented above demonstrates that the spin-spin term has
nothing to do with the turn-on of the spectral weight, as foreshadowed
by Mott\cite{mott}.  Namely, the gap in the spectrum at half-filling
is independent of ordering.  While a naive scaling analysis suggests
that the spin-spin interaction is indeed subdominant, we have been
unable to compute the $\beta$-function to show that a true
fixed point underlies the physics at half-filling.  Such a
computation stands as a true challenge for Mottness.

Nonetheless, antiferromagnetism with an ordering wave-vector of
$(\pi,\pi)$ can also be understood within this formalism.  Within
this theory, there is
a natural candidate for the antiferromagnetic order parameter, namely
$B_{ij}=\langle g_{ij}\varphi^\dagger_i
c_{i,\uparrow}c_{j,\downarrow}\rangle$. The vacuum
expectation value of this quantity is clearly non-zero as it is easily
obtained from a functional derivative of the partition function with
respect to $\gamma_{\vec p}$.  That this is the relevant order
parameter instead of the traditional one follows from the fact that the spin-spin interaction and all
higher-order operators contained in $|b|^2$ are at least proportional
to $a^4$ ($a$ the lattice constant) and hence are subdominant to the
composite terms.  Hence, a non-traditional order parameter must govern the
turn-on of antiferromagnetism.  We advocate that $B_{ij}$ characterizes the antiferromagnet that describes the strong-coupling limit of the
Hubbard model and as a consequence the insulating state of the cuprates.   An
antiferromagnet of this kind has no continuity with
the antiferromagnet at weak coupling because it is mediated by the
collective mode $\varphi$ or $\tilde\varphi$.  Hence, both the Mott
gap and subsequent antiferromagnetic order emerge from composite
excitations that have no counterpart in the original UV Lagrangian but
only become apparent in a proper low-energy theory in which the
high-energy degrees of freedom are explicitly integrated out. Away
from half-filling, a similar state of affairs obtains.

\section{Hole Doping: Experimental Consequences}

The charge $2e$ boson has much to tell us about the normal state of a
doped Mott insulator.  Here we compute the electron spectral function,
the specific heat, the thermal conductivity, the optical conductivity
as well as the dielectric function.  In each of these, the charge $2e$
boson produces a distinct signature that accounts for the anomalies of
the doped state of a Mott insulator. 

\subsection{Spectral Function: Pseudogap}

Since we have demonstrated that
${\cal L}_{\rm Mott}$ captures the strong-coupling physics of the Mott
insulating state, we focus on the evolution of this theory with
doping.  The lack of any gradient terms in the action with respect to
the bosonic fields and the absence of any bare dynamics associated
with $\varphi_i$ suggests that we can treat $\varphi_i$ as a
homogeneous field.  Further, since we are not interested in the
dynamics on the Mott scale, we treat $\varphi$ as a static
field.   Its sole role is to mix the subsectors which
differ in the number of doubly occupied sites. Consequently, our results are valid
provided that $\omega< U$ and $U\gg t$. Under these assumptions, the single-particle electron Green function
\beq
G(k,\omega)=-iFT \langle {Tc_i(t) c^\dagger_j(0)} \rangle,
\eeq
can be calculated rigorously in the path-integral formalism as
%\begin{widetext}
\beq
G(k,\omega)= -i FT \int [D\varphi_i^*] [D\varphi_i] \int [Dc_i^*]
[Dc_i]  c_i(t) c^*_j(0) \exp^{-\int {\cal L} [c,\varphi] dt},
\eeq
%\end{widetext}
where $FT$ refers to the Fourier transform and ${\bf T}$ is the
time-ordering operation. The explicit spin-spin term is not contained
in ${\cal L_{\rm Mott}}$.  This term will also be dropped in the doped case because even
in this limit, the spin-spin term is subdominant (in a naive continuum
limit sense) to the other interactions in ${\cal L}$.  This state of
affairs arises because spin-spin term in $b^\dagger b$ contains four
spatial derivatives, whereas the $\varphi^\dagger b$ term contains
only two.   As a
result, all of the physics we present below is
associated with the charge rather than the spin degrees of freedom.
The continuity of the analysis with that of the half-filled system
raises the question that perhaps a fixed point at half-filling
persists to finite doping as well in which only the fermion-boson
terms are relevant.  While our analysis is highly suggestive that
such a state of affairs obtains, the possible existence of such a fixed point
remains a conjecture as of this writing. 

To proceed, we will organize the calculation of $G(k,\omega)$ by first integrating out the fermions (holding $\varphi$ fixed)
%\begin{widetext}
\beq\label{geff}
G(k,\omega)= \int [D\varphi^*] [D\varphi] FT \left( {\int [Dc_i^*] [Dc_i]  c_i(t) c^*_j(0) \exp^{-\int L[c,\varphi] dt} }\right)
\eeq
%\end{widetext}
where now
%\begin{widetext}
\beq
{\cal L}&=&\sum_{i\sigma}(1-n_{i\bar\sigma}) c_{k\sigma}^* \dot c_{k\sigma} -\left( { 2\mu + \frac{s^2}{U} }\right)\varphi^*\varphi -\sum_{k\sigma}(g_t t \alpha_k+\mu) c_{k\sigma}^* c_{k\sigma}\nonumber\\
&+&s\varphi^*\sum_k (1-\frac{2t}{U}) c_{-k\downarrow}c_{k\uparrow}+c.c.
\eeq
%\end{widetext}
The effective Lagrangian can be diagonalized and written in terms of 
a collection of Bogoliubov quasiparticles\cite{ftm3}.  The remaining $\varphi$
integration can then be done numerically to obtain the spectral
function.  

\begin{figure}
\centering
\includegraphics[width=6.cm,angle=270]{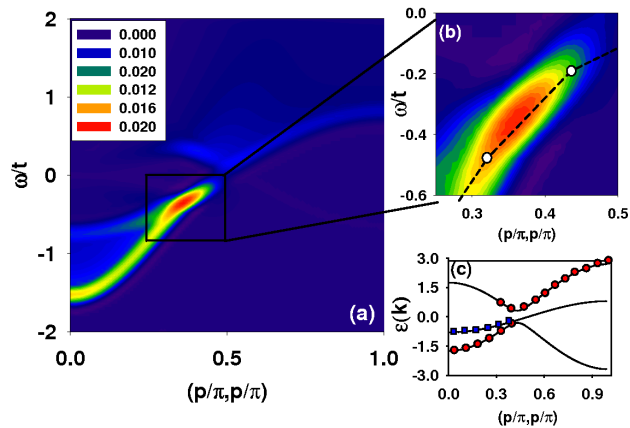}
\caption{(a) Spectral function for filling $n=0.9$ along the nodal direction.  The intensity is indicated by the color scheme.  (b) Location of the low and high energy kinks as indicated by the change in the slope of the electron dispersion.  (c) The energy bands that give rise to the bifurcation of the electron dispersion.}\label{specf1}
\end{figure}

\begin{figure}
\centering
\includegraphics[width=8.cm]{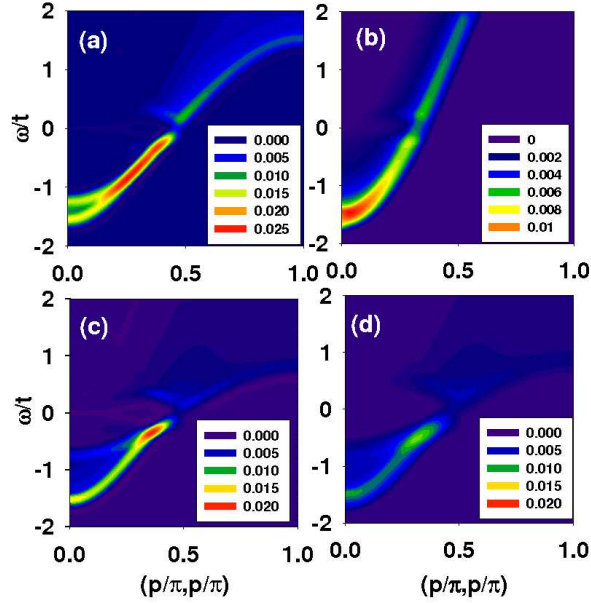}
\caption{Spectral function for two different fillings (a) $n=0.8$ and (b) $n=0.4$ along the nodal direction. The absence of a splitting in the electron dispersion at $n=0.4$ indicates the bifurcation ceases beyond a critical doping. The spectral functions for two different values of the on-site repulsion,
 (c)$U=10t$ and (d)$U=20t$ for $n=0.9$ reveals that the high-energy kink and the splitting of the electron dispersion have at best a weak dependence on $U$.  This indicates that this physics is set by the energy scale $t$ rather than $U$.}\label{specf2}
\end{figure}

The spectral functions in Figs. (\ref{specf1}) and (\ref{specf2}) exhibit four key features.  First, 
there is a low-energy kink in the electron dispersion that is
independent of doping.   The low-energy kink occurs at roughly $0.2t\approx 100 meV$.  By treating the mass term for the boson as a variable parameter, we verified
that the low-energy kink is determined by the bare mass.  In the effective low-energy theory, the bare mass is $t^2/U$.  This mass is independent of doping.
Experimentally,
the low-energy kink\cite{lenkink} does not change with doping.
Consequently, the charge $2e$ bosonic field provides a natural mechanism
for the kink that is distinct from the phonon schemes that have been proposed\cite{lenkink}.

Second,  a high-energy kink appears at roughly $0.5t\approx 250 meV$
which closely resembles the experimental kink at
$300meV$\cite{henkink}.  Cluster\cite{jkink} and exact diagonalization
methods\cite{pink} also find a high-energy kink. At sufficiently high doping (see
Figs. (\ref{specf2}a) and (\ref{specf2}b)), the high-energy kink
disappears.  Third, the electron dispersion bifurcates at the second
kink.  This is precisely the behaviour that is seen experimentally\cite{henkink}. 
The energy difference between  the two branches is maximum at
$(0,0)$ as is seen experimentally.  A computation of the spectral
function at $U=20t$ and $n=0.9$ reveals that the dispersion as well
the bifurcation still persist.  Further, the magnitude of the
splitting does not change, indicating that the energy scale for the
bifurcation and the maximum energy splitting are set by $t$ and not
$U$. The origin of the two branches is captured in
Fig. (\ref{specf1}c).  The two branches below the chemical potential
correspond to the standard band in the LHB (filled squares in
Fig. (\ref{specf1}c) on which $\varphi$ vanishes and a branch on which
$\varphi\ne 0$ (filled circles in Fig. (\ref{specf1}c). Simulations on
the Hubbard model clearly resolve either the low-energy
feature\cite{cluster1,cluster2,mg1} or the high-energy
kink\cite{jkink,pink}.  In the studies showing the hihg-energy kink,
the low-energy feature is not discernible\cite{jkink,pink}. What is
new here is that both features (but with drastically different
intensities as is seen experimentally) are captured.  The two
branches indicate that there are two local maxima in the integrand in
Eq. (\ref{geff}), a feature not captured by a saddle-point approximation. Above the chemical potential only one branch survives.  The split
electron dispersion below the chemical potential is consistent with
the composite nature of the electron operator dictated by
Eq. (\ref{cop}).  At low energies, the electron is a linear
superposition of two states, one the standard band in the LHB
described by excitations of the form,
$c_{i\sigma}^\dagger(1-n_{i\bar\sigma})$ and the other a composite
excitation consisting of a bound hole and the charge $2e$ boson,
$c_{i\bar\sigma}\varphi_i^\dagger$. The former contributes to the
static part of the spectral weight transfer ($2x$) while the new charge
$e$ excitation gives rise to the dynamical contribution to the spectral
weight transfer. Because the new charge $e$ state is strongly dependent on the
hopping, it should disperse as is evident from
Fig. (\ref{specf2})
and also confirmed experimentally\cite{henkink}.

\begin{figure}
\centering
\includegraphics[width=4.cm, angle=270]{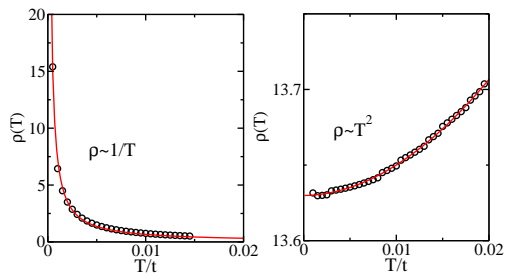}
\caption{(a)dc electrical resistivity as a function of temperature for $n=0.9$ (b) Setting the bosonic degree of freedom to zero kills the divergence of the resistivity as $T\rightarrow 0$. This suggests that it is the strong binding between between the fermionic and bosonic degrees of freedom that ultimately leads to the insulating behaviour in the normal state of a doped Mott insulator.}\label{dcr}
\end{figure}

The formation of the composite excitation,
$c_{i\bar\sigma}\varphi^\dagger$, is the new dynamical degree of
freedom in the doped theory.  This dynamical degree of freedom has no
counterpart in the UV scale.  Such a binding of a hole and the charge
$2e$ bosonic field leads to a pseudogap at the chemical
potential, as evidenced by the absence of spectral weight at the
chemical potential for both $n=0.9$ and $n=0.8$.  Non-zero spectral weight resides at the chemical potential
in the heavily overdoped regime, $n=0.4$, consistent with the
vanishing of the pseudogap beyond a critical doping away from
half-filling. Because the density of states vanishes at the chemical
potential,  the electrical resistivity diverges as $T\rightarrow 0$.
Such a divergence is shown in Fig. (\ref{dcr}a) and is consistent with
our previous calculations of the dc resistivity using a local
dynamical cluster method\cite{holeloc}.  In the absence of the boson (Fig. (\ref{dcr}b)), localization ceases.
Although this calculation does not constitute a proof, it is consistent with
localisation induced by the formation of the bound composite
excitation, $c_{i\bar\sigma}\varphi_i^\dagger$.  This state of affairs
obtains because the boson has no bare dynamics.  It may acquire
dynamics at $O(t^3/U^2)$ as can be seen by expanding the ${\cal M}$
matrix in Eq. (\ref{HIR}).  

Such bound-state formation lays plain how the strong
coupling regime of a doped Mott insulator depends on the 
dimensionality, the doping and the connectivity of the lattice.  As the
charge 2e boson is a local degree of freedom with no bare dynamics, an
analogy with bound state formation by a local potential is warranted.
It is well known that bound
state formation in $d\le 2$ obtains for an arbitrarily weak local
potential. For higher dimensions, a local potential
exceeding a threshold value is required for a bound state to form. That such a picture of the
bound-state formation applies here is supported by simulations on the
Hubbard model.  In $d=\infty$\cite{nopg} a pseudogap is absent, whereas
 a variety of strong-coupling cluster methods all yield a pseudogap\cite{pg1,pg2} without invoking
symmetry breaking on a $d=2$ square lattice in the vicinity of
half-filling. Since $L>2x$ is also a signature of a pseudogap (which
is mediated by bound-state formation), we
conclude that dynamical spectral weight transfer also depends on the
dimensionality of the lattice.  The absence of a pseudogap in
$d=\infty$ implies that must be some upper critical dimension above
which the interactions generated by the $t/U$ corrections in
Eq. (\ref{trans}) become irrelevant.  The precise nature of this fixed
point remains an open problem.

\begin{figure}
\centering
\includegraphics[width=4.5cm, angle=270]{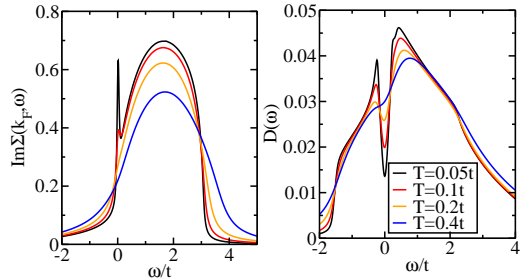}
\caption{The imaginary part of the self energy as the function of
 temperature for $n=0.7$. A peak is developed at $\omega=0$ at low
 temperature which is the signature of the opening of the
 pseudogap. The density of states explicitly showing the pseudogap is
shown in adjacent figure.}\label{self}
\end{figure}

A gap in the spectrum is possible only if the
single-particle Green function vanishes along some surface in momentum
space.  Along such a surface, the self-energy diverges.  The imaginary
part of the self energy at different temperatures is shown in
Fig. (\ref{self}). At low temperature ($T\leq t^2/U$),  the imaginary
part of the self-energy at the non-interacting Fermi surface develops
a peak at $\omega=0$.  At $T=0$, the peak leads to a divergence and
hence is consistent with the opening of a pseudogap. As we have pointed out earlier\cite{zeros}, a pseudogap is properly identified by a zero surface (the Luttinger surface) of the single-particle Green function. This zero surface is expected to preserve the Luttinger volume if the pseudogap lacks particle-hole symmetry as shown in the second of the figures in Fig. (\ref{self}).

\subsection{Mid-Infrared Band}

Naively, doped Mott insulators are expected to either have a 
far-infrared or an ultra-violet or upper-Hubbard-band scale
absorption. Hence, one of the true surprises in the optical response
of the cuprates is the  mid-infrared band
(MIB).    While many mechanisms have been proposed\cite{opt3}, no explanation has risen to the fore.
A hint as to the origin of this band is that the intensity in the MIB increases with doping at the expense of spectral weight at high energy and the energy scale for the peak in the MIB
is the hopping matrix element $t$.  Since the MIB arises from the
high-energy
scale, the current theory which accurately integrates
out the high energy degrees of freedom should capture this physics.  We work in the non-crossing approximation,
\beq\label{cond}
\sigma _{xx} (\omega ) &=& 2
\pi e^2
\int d^2 k\int d\omega '(2t\sin k_x )^2 \nonumber\\
&& \left(  -\frac{f(\omega ')-f(\omega'+\omega)}{\omega} \right)
A(\omega+\omega ',k)A(\omega',k),\nonumber
\eeq
to the Kubo formula for the
conductivity where $f(\omega)$ is the Fermi distribution function and
$A(\omega,k)$ is the spectral function.  In our treatment, the
vertex corrections arise solely from the interactions with the bosonic degrees of freedom.  Since the boson acquires dynamics only through electron motion and the
leading such term is $O(t^3/U^2)$, the treatment here should suffice to provide the leading behaviour of the optical conductivity.
\begin{figure}
\centering
\includegraphics[width=6.cm]{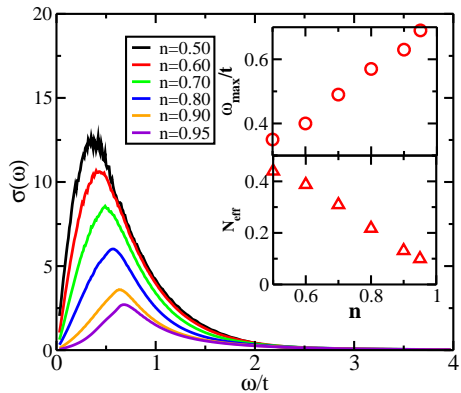}
\caption{Optical conductivity as a function of electron filling, $n$
 with the Drude part subtracted. The peak in the optical conductivity represents the mid-infrared band.  Its origin is mobile double occupancy in the lower-Hubbard band.  The insets show that the energy at which the MIB acquires its maximum value, $\omega_{\rm max}$ is an increasing function of electron filling.  Conversely, the integrated weight of the MIB decreases as the filling increases.  This decrease is compensated with an increased weight at high (upper-Hubbard band) energy scale. }\label{optcond1}
\end{figure}

\begin{figure}
\centering
\includegraphics[width=6.cm]{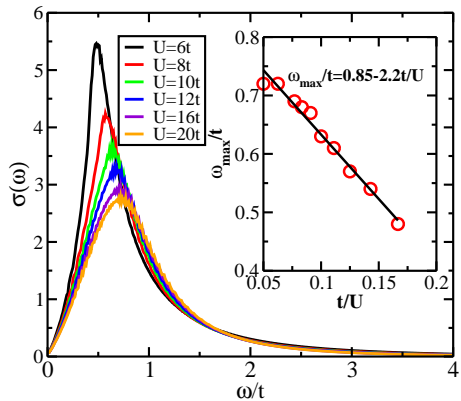}
\caption{Evolution of the optical conductivity for $n=0.9$ as $U$ is varied.
The inset shows the functional form that best describes $\omega_{\rm max}$.  The dominant energy scale is the hopping matrix element $t$ since $t/U$ for the cuprates is $O(1/10)$.}
\label{optcond2}
\end{figure}

Shown in Fig. (\ref{optcond1}) is the optical conductivity which peaks
at $\omega_{\rm max}\approx .5t$ forming the MIB. We have subtracted the
Drude weight at $\omega=0$ to focus sharply on the MIB. As the inset indicates, $\omega_{\rm max}$ is
an increasing function of the electron filling ($n$), whereas the integrated weight
\beq
N_{\rm eff}=\frac{2m^\ast}{\pi e^2}\int_0^{\Omega_c} \sigma(\omega)d\omega
\eeq
decreases.    However, $N_{\rm eff}$ does not vanish at half-filling indicating that the mechanism that causes the mid-IR is evident even in the Mott state.  We set the integration cutoff to $\Omega_c=2t=1/m^\ast$.
The magnitude and filling dependence of $\Omega_{\rm max}$ are all consistent with
that of the mid-infrared band in the optical conductivity in the
cuprates\cite{cooper,uchida1,opt1,opt2,opt3}.  We determined what sets
the scale for the MIB by studying its evolution as a function of $U$.
As is clear from Figure (\ref{optcond2}), $\omega_{\rm max}$ is set essentially by the hopping matrix
element $t$ and depends only weakly on $J$. The physical processes that determine this physics are determined by the coupled boson-Fermi terms in the low-energy
theory.  The $\varphi_i^\dagger c_{i\uparrow}c_{i\downarrow}$ term has
a coupling constant of $t$ whereas the $\varphi_i^\dagger b_i$ scales
as $t^2/U$.
Together, both terms give rise to a MIB band that scales as $\omega_{\rm max}/t=0.8-2.21t/U$ (see inset of Fig. (\ref{optcond2})). Since $t/U\approx O(.1)$
for the cuprates, the first term dominates and the MIB is determined
predominantly by the hopping matrix element $t$. Within the
interpretation that $\varphi$ represents a bound state between a
doubly occupied site and a hole, second order perturbation theory with
the $\varphi_i^\dagger b_i$ term mediates the process shown in
Fig. (\ref{dos}).  It is the resonance between these two states that
results in the mid-IR band.  Interestingly, this resonance persists even at half-filling and hence the non-vanishing of $N_{\rm eff}$ at half-filling is not evidence that the cuprates are not doped Mott insulators as has been recently claimed\cite{millis}. 

As the physics in Fig. (\ref{dos}) is not present in projective
models which prohibit double occupancy in the Hubbard basis (not simply
the transformed fermion basis of the t-J model), it is instructive to see what calculations of the optical conductivity in the $t-J$ model
reveal.  All existing calculations\cite{opt2,haulek,prelovsek,zaanen} on the $t-J$ model find that the MIB scales as $J$.  In some of these calculations, superconductivity
is needed to induce an MIB\cite{haulek} also at an energy scale of
$J$. In others, phonons are invoked to overcome the failure of the
hard-projected t-J model to yield a mid-infrared band. Experimentally\cite{cooper,uchida1,opt3}, it is clear that the MIB is set by the $t$ scale
rather than $J$.  In fact, since the MIB grows at the expense of spectral weight in the upper-Hubbard band, it is not surprising that the t-J model cannot describe
this physics as first pointed out by
Uchida, et al.\cite{uchida1}.   The physical mechanism we have
identified here, Fig. (\ref{dos}) clearly derives from the high energy scale,
has the correct energy dependence, and hence satisfies the key
experimental constraints on the origin of the MIB. Since the physics in Fig. (\ref{dos}) is crucial to the mid-IR, it is not surprising that single-site analysis\cite{millis} fail to obtain a non-zero intercept in the extreme Mott limit. The non-zero intercept of $N_{\rm eff}$ is a consequence of Mottness and appears to be seen experimentally in a wide range of cuprates\cite{cooper,int1,int2,int3}.

\subsection{Dielectric function: Experimental Prediction}

We have shown thus far that there are two branches in the electronic
spectral function below the chemical potential.  Such physics is
explained by the dynamical formation of a new
composite excitation, representing a bound state,  consisting of a
bound hole and a charge $2e$ boson, $\varphi_i^\dagger
c_{i\bar\sigma}$.  We demonstrated that for the MIB in the optical
conductivity such an excitation also appears.  Such 
composite charge excitations should show up in response
functions which are sensitive to all the charge degrees of freedom, for example, the energy loss function, $\Im
1/\epsilon(\omega,\vec q)$, where $\epsilon(\omega,\vec q)$ is the
dielectric function.  We show here that this is the case.

\begin{figure}
\centering
\includegraphics[width=6.cm]{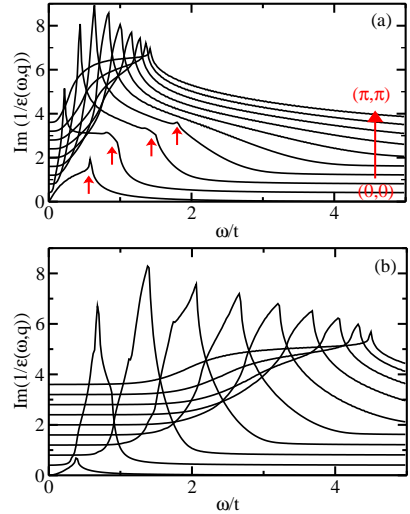}
\caption{The dielectric function, $-\Im 1/\epsilon(\omega,\vec q)$
for $\vec q$ along the diagonal direction is shown for (a) $n=0.9$
and (b) $n=0.6$. Note only the broad feature indicated by the red arrow at $n=0.9$ persists at
$n=0.6$.}
\label{dielect}
\end{figure}

To this end, we calculate the inverse dielectric function,
\beq
\Im \frac{1}{\epsilon(\omega,\vec q)} &=& \pi \frac{U}{v}\sum_p \int d\omega'
(f(\omega')-f(\omega+\omega'))
\times A(\omega+\omega',\vec p+\vec q) A(\omega',\vec p),\nonumber \eeq
using the non-crossing approximation discussed earlier. Our results
are shown in Fig.(\ref{dielect}) for $n=0.9$ and $n=0.6$ for $\vec
q$ along the diagonal. Two features are distinct. First, there is a
broad band (red arrow in Fig. (\ref{dielect})) with the width of order $t$ that disperses with $\vec q$
for both doping levels. It is simply the particle-hole continuum
which arises from the renormalized bare electron band. The band width is doping
dependent as a result of the renormalization of the band with
doping.  More strikingly,  for $n=0.9$, a sharp peak exists at
$\omega/t\approx .2t$.  It disperses with $q$, terminating when
$\vec q\rightarrow (\pi, \pi)$.   Physically, the sharp peak
represents a quasiparticle excitation of the composite object,
$\varphi_i^\dagger c_{i\bar\sigma}$, the charge 2e boson and a hole.
Therefore, we predict that if this new composite charge excitation,
$\varphi_i^\dagger c_{i\bar\sigma}$, is a real physical entity, as
it seems to be, it will give rise to a sharp peak in addition to the
particle-hole continuum in the inverse dielectric function.  Since
this function has not been measured at present, our work here
represents a prediction.  Electron-energy loss spectroscopy can be
used to measure the inverse dielectric function. Our key prediction
is that momentum-dependent scattering should reveal a sharp peak
that appears at low energy in a doped Mott insulator.  We have
checked numerically the weight under the peaks in the inverse
dielectric function and the sharp peak is important.  Hence, the new
charge $e$ particle we have identified here should be experimentally
observable.  The two dispersing particle-hole features
found here are distinct from a similar feature in stripe
models\cite{zaanen2}. 
In
such models the second branch\cite{zaanen2} has vanishing weight and
whereas
in the
current theory both features are of unit weight.

\subsection{Heat conductivity and heat capacity}

\begin{figure}
\centering
\includegraphics[width=4.cm, angle=270]{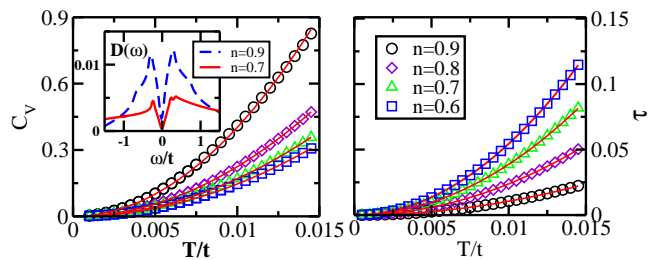}
\caption{(a) Heat capacity, $C_V$, and (b) thermal conductivity, $\tau$, calculated at $n=0.9$.  The solid lines are a fit to $T^2$. Insert: Density of states for $U=10t$ are evaluated at $n=0.9$ and $n=0.7$ respectively.}
\label{thermal}
\end{figure}

Loram and collaborators\cite{loram} have shown from their extensive
measurements that the heat capacity in the cuprates in the normal
state scales as $T^2$.  It is a trivial exercise to show that such a
temperature dependence requires a  V-shaped gap density of states as a
function of energy.  The slope of the density of states in the
vicinity of the chemical potential determines the
coefficient of the $T^2$ term.   Because the slope of the density of
states decreases as the pseudogap closes, the magnitude of the $T^2$
term should diminish as the doping increases.  As we showed in the
previous section, the boson creates a pseudogap.  The energy
dependence of the gap is shown in the inset of Fig. (\ref{thermal}).
A linear dependence on energy is apparent.  We calculated the heat capacity shown in Fig.(\ref{thermal}a) via the relationship $C_v=\frac{d \bar E}{dt}$, where the internal energy, $\bar E$, is
\beq
\bar E =\int d\omega D(\omega) \omega f(\omega)
\eeq
and $D(\omega)=\sum_{\vec k}A(\omega, \vec k)$.  As expected, the
temperature dependence is quadratic in the doping regime where the
pseudogap is present as is seen experimentally\cite{loram}.  As it is
the boson that underlies the pseudogap, it is the efficient cause of
the $T^2$ dependence of the heat capacity.  In our theory, the steeper slope occurs at smaller doping which gives
rise to the largest heat capacity at half filling. This doping
dependence of the heat capacity seems to contradict the experimental
observations\cite{loram}. A key in determining the magnitude of the
heat capacity is the spin degrees of freedom.  As we have focused
entirely on the bosonic degree of freedom and not on the contribution
from the spin-spin interaction terms, we have over-estimated the
kinetic energy. Such terms do not affect the pseudogap
found here though they do change the doping dependence\cite{prelovsek}.  From Eq. (\ref{cop}) it is clear that the spin-spin terms
renormalize the standard fermionic branch in the lower-Hubbard band
leaving the new state mediated by $\varphi_i$ untouched.  

Additionally, the thermal conductivity, $\tau(T)$, can be calculated using the Kubo formula in non-crossing approximation,
\beq
\tau(T)=\frac{e}{4k_B T} \sum_{\vec k}\int \frac{d\omega}{2\pi} (v_{\vec k}^x)^2 \omega^2
\left(-\frac{\partial f(\omega)}{\partial \omega} \right) A(\vec
k,\omega)^2.\nonumber
\eeq
The thermal conductivity shown in Fig.(\ref{thermal}) scales as $T^2$ which is identical to that of the heat capacity.  However, the system exhibits a larger thermal conductivity as the doping increases in contrast to the heat capacity which is decreasing as the doping increases. Physically, this signifies that the carriers are more mobile as the doping increases.

%In order to compute the thermal conductivity using the Kubo formula, we
% define several current-current correlation functions:
%\beq
%L^{11}&=&-\frac{1}{\beta}\int_0^\infty dt e^{-st}\int_0^\beta d\beta' Tr[\rho_0 j(-t-i\beta') j]\\
%L^{12}&=&-\frac{1}{\beta}\int_0^\infty dt e^{-st}\int_0^\beta d\beta' Tr[\rho_0 J_Q(-t-i\beta') j]\\
%L^{22}&=&-\frac{1}{\beta}\int_0^\infty dt e^{-st}\int_0^\beta d\beta' Tr[\rho_0 J_Q(-t-i\beta') J_Q]
%\eeq
%where $J_Q$ is the heat current and $j$ is the electric current. The Kubo formula for the thermal conductivity
%\beq
%\label{eq:thermal-cond}
%\tau(T)&=&\left( \frac{L^{22}}{T^2} - \frac{(L^{12})^2}{L^{11}} \right)
%\eeq
%reduces to
%\beq
%\tau(T)=\frac{e}{4k_B T} \sum_{\vec k}\int {d\omega}{2\pi} (v_{\vec k}^x)^2 \omega^2
%\left(-\frac{\partial f(\omega)}{\partial \omega} \right) A(\vec k,\omega)^2
%\eeq
%in the non-crossing approximation.
%\beq
%\tau=\frac{1}{3}C_V v_s l_s
%\eeq
%$v_s$ is the sound velocity

\subsection{T-linear Resistivity}

A key theme of this review is that the normal state of doped Mott
insulators is dominated by dynamical degrees of freedom that could
not have been deduced from the UV physics.  Further, as stated in the
introduction, the correct theory of the pseudogap phase should also
explain the $T-$linear resistivity.  The standard explanation\cite{tlin} attributes $T-$ linear resistivity to quantum criticality. However, one of us has recently shown\cite{tlin} that under three general assumptions, 1) one-parameter scaling, 2) the critical degrees of freedom carry the current and 3) charge is conserved, the resistivity in the quantum critical regime takes the universal form,
\begin{eqnarray}\label{dclimit}
\sigma(\omega=0)=\frac{Q^2}{\hbar}\;\Sigma(0)\;\left(\frac{k_BT}{\hbar
   c}\right)^{(d-2)/z} .
\end{eqnarray}
As a result, quantum criticality in its present form yields
$T-$linear resistivity (for d=3) only if the dynamical exponent
satisfies the unphysical constraint $z<0$.  The remedy here might be
three-fold:  1)  some other yet-unknown phenomenon  is responsible for
$T-$linear resistivity, 2) the charge carriers are non-critical, or 3)
the single-parameter scaling hypothesis must be relaxed. 

\begin{figure}
\centering
\includegraphics[width=7.5cm]{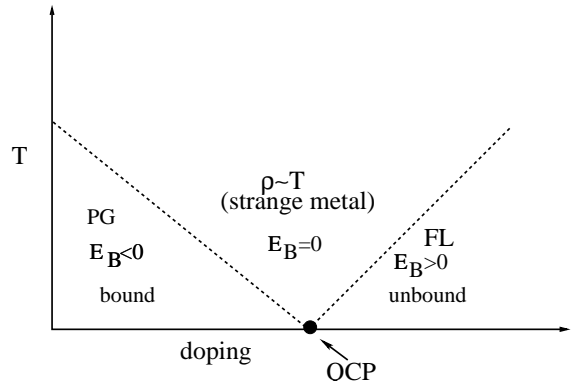}
\caption{Proposed phase diagram for the binding of the holes and bosons that result in the formation of the pseudogap phase.  Once the binding energy vanishes, the energy to excite a boson vanishes.  In the critical regime, the dominant scattering mechanism is still due to the interaction with the boson. T-linear resistivity results anytime $T>\omega_b$, where $\omega_b$ is the energy to excite a boson.  To the right of the quantum critical regime (QCP), the boson is irrelevant and scattering is dominated by electron-electron interactions indicative of a Fermi liquid. The QCP signifies the end of the binding of fermi and bosonic degrees of freedom that result in the pseudogap phase.}
\label{tlin}
\end{figure}

The new dynamical degree of freedom we have identified here fits the
bill and provides the added ingredient to explain $T-$linear resistivity.
While none of the calculations presented here is sufficient to account
for the confined dynamics of $\varphi_i$, the
formation of the pseudogap, the divergence of the electrical
resistivity, the $\varphi_i^\dagger c_{i\bar\sigma}$ feature in the
electron operator, and the new feature in the dielectric function all
point in this direction.  Consequently, we assume that $\varphi_i^\dagger
c_{i\bar\sigma}$ forms a bound state and the binding energy is $E_B$. As a bound
state, $E_B<0$, where energies are measured relative to the chemical
potential.  Upon increased hole doping, the chemical potential
decreases.  Beyond a critical doping, the chemical potential,
crosses the energy of the bound state.  At the critical value of the
doping where $E_B=0$, the energy to excite a boson vanishes. The
critical region is dominated by electron-boson scattering. In
metals, it is well-known\cite{bass} that above the Debye
temperature, the resistivity arising from electron-phonon scattering
is linear in temperature.  We make a direct analogy here with the electrons
scattering off phonons in a metal.  Once the boson unbinds, we assume its
dynamics is purely classical.  Since the energy to create a boson
vanishes at criticality as shown in Fig. (\ref{tlin}), T-linear
resistivity obtains.
Namely, in the critical region, the energy to create a boson
vanishes as shown in Fig. (\ref{tlin}) and hence the resistivity
arising from electron-boson scattering should be linear in
temperature. This mechanism is robust (assuming the unbound boson has classical dynamics)  as it relies solely on the
vanishing of the boson energy at criticality and not on the form of
the coupling.  To the right of the quantum critical point, standard
electron-electron interactions dominate and Fermi liquid behaviour
obtains.  In this scenario, the quantum critical point coincides
with the termination of the pseudogap phase, or equivalently with
the unbinding of the bosonic degrees of freedom. Since it is the bound state of the boson that creates the new charge e state giving rise to $L/n_h>1$ and this state is generated as a result of dynamical spectral weight transfer, the $T*$ line defines the temperature below which dynamical spectral weight transfer contributes to the low-energy spectral weight.  Consequently, the mechanism proposed here is experimentally testable.  Simply repeat the x-ray K-edge experiments presented in Fig. (\ref{fy}) below and above the $T*$ line.  Above $T^*$ the integrated weight should be $2x$ whereas below it should exceed $2x$. 

\subsection{Towards Superconductivity}

\begin{figure}
\centering
\includegraphics[width=9.0cm,angle=90]{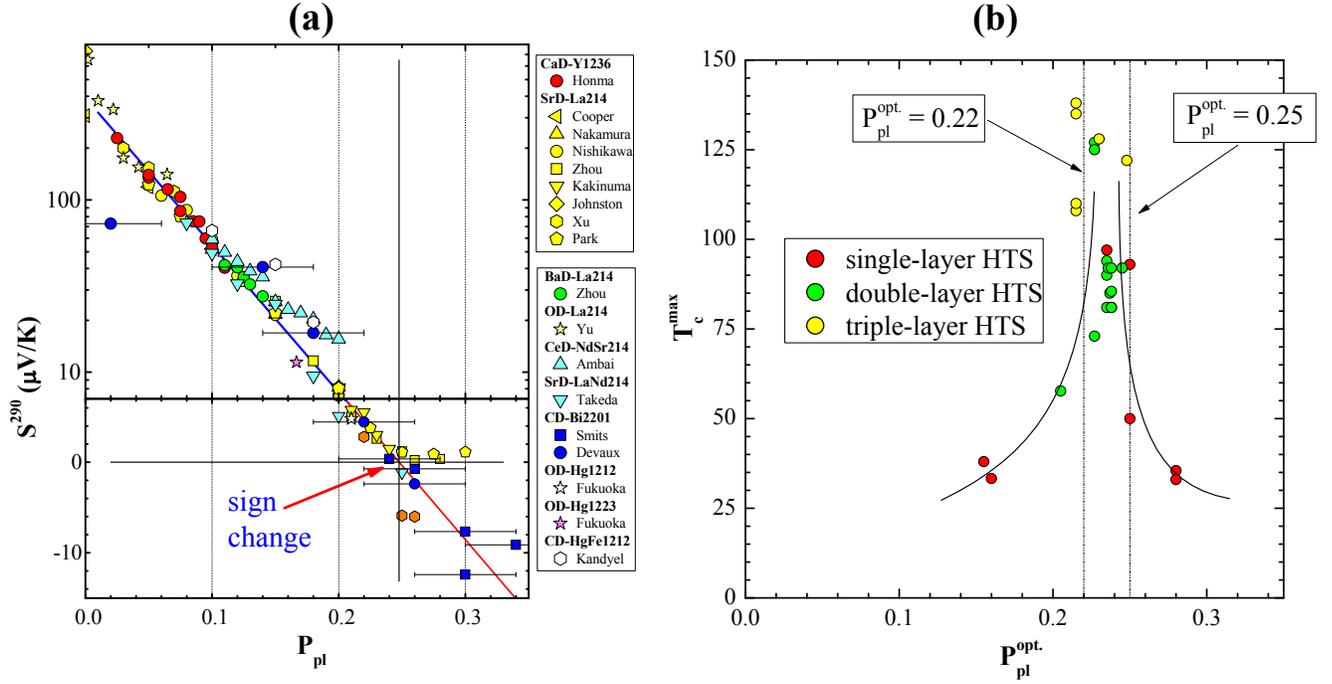}
\caption{ a) Universal behavior of  the thermoelectric power\cite{vanishingtp} (290K) as a
 function of planar hole density ($P_{\rm pl}$), for various families
of hole-doped
 cuprates.   All exhibit a sign change at $P_{\rm pl}=0.23$. Above the solid-bold horizontal line, the thermpower obeys the
 functional form, $S^{290}(P_{\rm pl})=392\exp(-19.7P_{\rm pl})$ for
 $0.01<P_{\rm pl}<0.21$.  Below the solid-bold horizontal line,
 $S^{290}(P_{pl})=40.47-163.4P_{\rm pl}$ for $0.21<P_{\rm pl}<0.34$.
 These functional forms were used\cite{vanishingtp} to determine the hole doping levels
 for all the cuprates rather than the widely used empirical formula\cite{thermoboys}
 $1-T_c/T_c^{\rm max}=82.6(x-0.16)^2$  which artificially fixes the
 optimal doping level of all cuprates to be $0.16$.  The thermopower
 scale is unbiased in this regard and has been shown\cite{vanishingtp} to corroroborate
 independent measures of the doping level even in Y123 and Tl-2201 in
 which it is the oxygen content that determines the doping level. b)
 Maximum transition temperature as a function of the planar hole
 density using the thermopower scale to determine the doping level.  Except for three single-layer materials, the vanishing of
 the thermopower coincides with the doping level
 at which the transition temperature is maximized.  }
\label{thermo}
\end{figure}

Our emphasis thus far has been on identifying a unifying principle for
the normal state of the cuprates.  As we
have seen, strong correlations mediate new composite excitations made
partly out of the emergent charge $2e$ boson that results by exactly
integrating out the high-energy scale in the Hubbard model. An
important question concerns the relevance of the physics we have
identified here to the superconducting phase.  Equivalently, what
role, if any,
does dynamical spectral weight transfer play in the superconducting
state?   We answer this question by focusing on a correlate of
high-temperature superconductivity. As the phase diagram indicates,
the superconducting region is roughly dome-shaped.  Why
superconductivity peaks at a particular doping level is
not known.  To offer some insight into this puzzle, we focus on an experimental
quantity which exhibits an abrupt sign change near optimal doping.
As shown in Fig. (\ref{thermo}), at a doping level corresponding to the highest superconducting
transition temperature for a wide range of cuprates, the thermopower
vanishes\cite{vanishingtp}.  
Consequently,
the sign change of $S$ occurs at the doping value defining the top of
the ``dome''.   While this might be an accident, the fact
that the thermopower vanishes at the same doping
level for most cuprates indicates that the reason might have something to do
with the superconducting mechanism. 

The thermopower measures the thermoelectric voltage induced across a
material in response to an applied temperature gradient. Microscopically, the thermopower is a measure of
the entropy per charge carrier. Further, it reveals the nature of the
dominant charge carriers, being positive for holes and negative for
electrons.  Should the entropy per carrier be identical for particles
and holes, the thermopower vanishes.  Consider hole-doping a Mott insulator.  Because transport obtains in the lower-Hubbard band, naively a vanishing of the
thermopower is expected whenever the number of states above and below
the chemical potential is equal.  In the atomic limit, this
corresponds to the condition $2x=1-x$, the solution of which is
$x_{\rm crit}=1/3$.  This result is corroborated by the large $U$ limit of the thermopower,
\beq
S=-\frac{k_B}{e}\ln\frac{2x}{1-x},
\eeq
computed by Beni\cite{beni} roughly 20 years before spectral weight transfer was
discovered.
Since  $2x$ and $1-x$ are the exact values for the electron addition
and removal states, respectively, in the atomic limit, it is
easy to see that the logarithm is precisely the entropy per carrier.
The logarithm vanishes at $x_{\rm crit}=1/3$ which is the exact particle-hole symmetric
condition for the LHB in the atomic limit.  Finite $t/U$ corrections
will increase $L$ and as a result decrease $x_{\rm crit}$. However, this is not all.  The spectral function is strongly momentum dependent when $t/U\ne 0$. As a result, strict particle-hole symmetry is not needed to make the thermopower (or even the Hall coefficient) vanish as can be seen directly from the exact\cite{shila} expression. Both of these effects conspire\cite{shila} to move the doping level at which the thermopower vanishes significantly below the atomic limit of $x=1/3$.  As this change is made entirely from the $t/U$
corrections to the thermopower, it is the dynamical spectral weight
transfer that is ultimately responsible for the precise value of
doping at which the thermopower vanishes in the cuprates. Consequently,
the dynamical spectral weight transfer plays a role in maximizing $T_c$.
 As a result, superconductivity in the
cuprates is determined fundamentally by the mixing between the high
and low energy scales in a doped Mott insulator.  The collective
degree of freedom $\varphi$ which results from the spectral weight
transfer is consequently central to the superconducting mechanism.
Optical conductivity
experiments\cite{mkhe,bontempshe,vmhe} certainly have shown this plainly
 that the onset of superconductivity results in a decrease in
the spectral weight in the UHB.  It would seem then that the ultimate solution to superconductivity hinges on the precise dynamics of the collective charge $2e$ boson that
we have shown to exist in the exact low-energy theory of a doped
Mott insulator.

\ack This research was supported in part by the NSF
DMR-0605769, P. H. Hor and T. Honma for the use of their thermopower data and S. Chakraborty his characteristically level-headed remarks. P. Phillips thanks the Max-Planck Institute in Dresden for their hospitality during the writing of this article.


\begin{thebibliography}{99}
\bibitem{mott} Mott N F, 1949  The Basis of the electron theory of
 metals, with special reference to transition metals Proc. Phys. Soc. London, Series A {\bf 62},
 416-422.

\bibitem{stechel}Hybertsen M S, Stechel E B, Schluter M, and Jennison D R 1990
Renormalization from density-functional theory to strong-coupling
models for electronic states in Cu-O material Phys. Rev. B {\bf 41}
11068 - 11072.

\bibitem{hubbard} Hubbard J. 1964 Electron correlations in narrow
 bands. III. An Improved Solution Proc. Roy. Soc. London, Ser. A {\bf
   281} 401-419.

\bibitem{harris} Harris A B and Lange R V 1967 Single-Particle Excitations in Narrow Energy Bands
Phys. Rev. {\bf 157} 295-314.

\bibitem{fulde}Kaplan T A, Horsch P, and Fulde P 1982 Close Relation
  between localized-electron magnetism and the paramagnetic wave
  function of completely itinerant elecrons Phys. Rev. Lett. {\bf 49} 889-892.

\bibitem{castellani}Castellani C, Di Castro C, Feinberg D, and Ranninger J 1979
New Model Hamiltonian for the Metal-Insulator Transition Phys. Rev.
Lett. {\bf 43} 1957 - 1960.

\bibitem{alloul} Alloul H, Ohno T, and Mendels P 1989
89Y NMR evidence for a fermi-liquid behavior in
YBa$_2$Cu$_3$O$_{6+x}$ Phys. Rev. Lett. {\bf 63} 1700-1703.

\bibitem{norman} Norman M R, Ding H, Randeria M, Campuzano J C, Yokoya T, Takeuchi T, Takahashi T, Mochiku T,
Kadowaki K, Guptasarma P and Hinks D G 1998  Destruction of the
Fermi surface in underdoped high-Tc superconductors Nature (London)
{\bf 392} 157-160.

\bibitem{ftm1}Leigh R G,  Phillips P, and Choy T P 2007 Hidden charge 2e
 boson in doped Mott insulators 
Phys. Rev. Lett. {\bf 99} 046404/1-4.

\bibitem{ftm2} Choy T P, Leigh R G,  Phillips P, and Powell P D. 2008
 Exact integration of the high-energy scale in a doped Mott insulator
Phys. Rev. B {\bf 77}, 014512/1-12.

\bibitem{ftm3} Choy T P, Leigh R G,  Phillips P, 2008 Charge 2e boson:
 Experimental consequences for doped Mott insulators, Phys. Rev. B
 {\bf 77}, 104524/1-9.

\bibitem{Kanigel} Kanigel A, Norman M R, Randeria M, Chatterjee U, Souma S, Kaminski A,
Fretwell H M, Rosenkranz S, Shi M, Sato T, Takahashi T, Li Z Z,
Raffy H, Kadowaki K, Hinks D, Ozyuzer L, Campuzano J C 2006
Evolution of the pseudogap from Fermi arcs to the nodal liquid Nat.
Phys. 2 447-451.

\bibitem{timusk} Timusk T and  Statt B 1999 The pseudogap in high-temperature superconductors: an experimental survey
Rep. Prog. Phys. {\bf 62} 61-122.

\bibitem{batlogg}For a review, see Batlogg B, Takagi H, Kao H L, and Kwo J,
Electronic Properties of High-$T_c$ Superconductors, edited by
Kuzmany H, Mehring M and Fink J, Springer-Verlag, Berlin, 1993 pp.
5-12.

\bibitem{ando} Ando Y, Komiya S, Segawa K, Ono S, and Kurita Y 2004
Electronic Phase Diagram of High-Tc Cuprate Superconductors from a
Mapping of the In-Plane Resistivity Curvature Phys. Rev. Lett. {\bf
93} 267001/1-4.

\bibitem{stripes} Kivelson S A, Fradkin E, Emery V J 1998  Electronic liquid-crystal phases of a doped Mott insulator
Nature {\bf 393} 550-553.


\bibitem{ddw} Chakravarty S, Laughlin R B,  Morr Dirk K and Chetan
Nayak 2001 Hidden order in the cuprates Phys. Rev. B {\bf 63}
094503/1-10.

\bibitem{rvb} Anderson P W 1987
The Resonating Valence Bond State in La2CuO4 and Superconductivity
Science {\bf 235} 1196 - 1198.

\bibitem{pfp} Emery V J and Kivelson S A 1995
Importance of phase fluctuations in superconductors with small
superfluid density Nature {\bf 374} 434-437.

\bibitem{inco1} Randeria M, Trivedi N, Moreo A and Scalettar R T
1992 Pairing and spin gap in the normal state of short coherence
length superconductors Phys. Rev. Lett. {\bf 69} 2001-2004.

\bibitem{inco2} Ranninger J, Robin J M and Eschrig M
1995 Superfluid Precursor Effects in a Model of Hybridized Bosons
and Fermions Phys. Rev. Lett. {\bf 74} 4027-4030.

\bibitem{inco3} Franz M and Tesanovic Z 2001
Algebraic Fermi liquid from phase fluctuations: Topological
fermions, vortex berryons and QED3 Theory of Cuprate
Superconductors Phys. Rev. Lett. {\bf 87} 257003/1-4.

\bibitem{nernst} Xu Z A, Ong N P, Wang Y, Kakeshita T and Uchida S 2000
Vortex-like excitations and the onset of superconducting phase
fluctuation in underdoped La2-xSrxCuO4 Nature {\bf 406} 486-488.

\bibitem{inco4} Kanigel A, Chatterjee U, Randeria M, Norman M R,
Koren G, Kadowaki K, and Campuzano J C
2008 Evidence for pairing above Tc from the dispersion in the
pseudogap phase of cuprates arXiv:0803.3052.

\bibitem{stripes1} Tranquada J M, Woo H, Perring T G, Goka H, Gu G D, Xu G, Fujita M and Yamada K
2004  Quantum magnetic excitations from stripes in copper oxide
superconductors Nature {\bf 429}, 534-538.

\bibitem{stripes2} Zaanen J and Gunnarsson O 1989
Charged magnetic domain lines and the magnetism of high-Tc oxides
Phys. Rev. B {\bf 40} 7391-7394.


\bibitem{stripes3}Abbamonte P, Rusydi A, Smadici S, Gu G D,
Sawatzky G A, and Feng D L 2005 Spatially modulated 'Mottness' in
La$_{2-x}$Ba$_x$CuO$_4$ Nat. Phys. {\bf 1} 155-158.


\bibitem{stripes4}Fink J, Schierle E, Weschke E, Geck J,
Hawthorn D, Wadati H, Hu H H, Durr H A, Wizent N, Buchner B, and
Sawatzky G A 2008 Charge order in La$_{1.8-x}$Eu$_{0.2}$Sr$_x$CuO$_4$
studied by resonant soft X-ray diffraction arXiv:0805.4352.


\bibitem{stripes5}Pasupathy A N, Pushp A, Gomes K K, Parker C V, Wen J, Xu Z, Gu G, Ono S, Ando Y, and Yazdani
A 2008 Electronic Origin of the Inhomogeneous Pairing Interaction in
the High-Tc Superconductor Bi$_2$Sr$_2$CaCu$_2$O$_{8+\delta}$
Science {\bf 320} 196-201.

\bibitem{trsb1}  Jing X, Elizabeth S, Deutscher G, Kivelson S A, Bonn D A,
Hardy W N, Liang R, Siemons W, Koster G, Fejer M M, Kapitulnik A.
2008 Polar Kerr-Effect Measurements of the High-Temperature
YBa2Cu3O6+x Superconductor: Evidence for Broken Symmetry near the
Pseudogap Temperature 2008 Phys. Rev. Lett {\bf 100} 127002.

\bibitem{trsb2} Kaminski A, Fretwell H M, Campuzano J C, Li Z,
Raffy H, Cullen W G, You H, Olson C G, Varma C M and H\"ochet H 2002
Spontaneous breaking of time-reversal symmetry in the pseudogap state of a high-Tc superconductor
Nature {\bf 416} 610-613.

\bibitem{trsb3} Simon M E and Varma C M 2002
Detection and Implications of a Time-Reversal Breaking State in
Underdoped Cuprates Phys. Rev. Lett. {\bf 89} 247003/1-4.

\bibitem{tsrb4} Fauque B, Sidis Y, Hinkov V, Pailhe S, Lin C T,
Chaud X, and Bourges P 2006 Resonant Magnetic Excitations at High
Energy in Superconducting YBa$_2$Cu$_3$O$_{6.85}$ Phys. Rev. Lett. {\bf
96} 197001/1-4.

\bibitem{qoscill} Leyraud N D, Proust C, LeBoeuf D, Levallois J,
Bonnemaison J B, Liang R, Bonn D A, Hardy W N, and Taillefer L 2007
Quantum oscillations and the Fermi surface in an underdoped high-Tc
superconductor Nature {\bf 447} 565-568.


\bibitem{raffy}  Konstantinovic Z, Li Z Z, and Raffy H 2001
Evolution of the resistivity of single-layer
Bi$_2$Sr$_{1.6}$La$_{0.4}$CuO$_y$ thin films with doping and phase
diagram Physica C {\bf 351} 163-168.

\bibitem{cooper} Cooper S L, Thomas G A, Orenstein J, Rapkine D H, Millis A J,
Cheong S W, Cooper A S and Fisk Z 1990 Growth of the optical
conductivity in the Cu-O planes Phys. Rev. B {\bf 41} 11605-11608;
Cooper S L, Reznik D, Kotz A, Karlow M A, Liu R, Klein M V, Lee W C,
Giapintzakis J, Ginsberg D M, Veal B W and Paulikas S P 1993 Optical
studies of the a-, b-, and c-axis charge dynamics in
YBa$_2$Cu$_3$O$_{6+x}$ Phys. Rev. B {\bf 47} 8233-8248.

\bibitem{uchida1} Uchida S, Ido T and Takagi H, Arima T, Tokura Y and Tajima
S 1991 Optical spectra of La$_{2-x}$Sr$_x$CuO$_4$: Effect of carrier doping on
the electronic structure of the CuO2 plane Phys. Rev. B {\bf 43}
7942-7954.

\bibitem{opt0}van der Marel D, Molegraaf H J A, Zaanen J, Nussinov Z, Carbone F, Damascelli A, Eisaki H, \
Greven M, Kes P H and Li M 2003  Quantum critical behaviour in a
high-Tc superconductor Nature {\bf 425} 271-274.

\bibitem{opt1} Moore S W, Graybeal J M, Tanner D B, Sarrao J, and
Fisk Z 2002  Optical properties of La$_2$Cu$_{1-x}$Li$_x$O$_4$ in the mid-infrared
Phys. Rev. B {\bf 66} 060509/1-4.

\bibitem{opt2}Bouvier J, Bontemps N, Gabay M, Nanot M and Queyroux F 1992
Infrared reflectivity versus doping in YBa$_2$Cu$_3$O$_{6+x}$ and
Nd$_{1+y}$Ba$_{2-y}$Cu$_3$O$_{6+x}$ ceramics: Relationship with the
t-J model Phys. Rev. B {\bf 45} 8065-8073.

\bibitem{opt3} Lee Y S, Segawa K, Li Z Q, Padilla W J, Dumm M, Dordevic S V,
Homes C C, Ando Y, and Basov D N 2005 Electrodynamics of the nodal
metal state in weakly doped high-Tc cuprates , Phys. Rev. B {\bf 72}
054529/1-13.

\bibitem{polchinski} Polchinski J 1992 Effective Field Theory and the Fermi Surface hep-th/9210046.

\bibitem{shankar}Benfatto G and Gallavotti G 1990
Perturbation theory of the Fermi surface in a quantum liquid: A
general quasiparticle formalism and one-dimensional systems J. Stat.
Phys. {\bf 59} 541-664.

\bibitem{others} Benfatto G and Gallavotti G 1990
Renormalization-group approach to the theory of the Fermi surface
Phys. Rev. B {\bf 42} 9967-9972.

\bibitem{others2} Shankar R 1991 Renormalization
group for interacting fermions in $d > 1$ Physica {\bf A177} 530-536;
Volovik G E 2007 ``Quantum Analogues: From Phase Transitions to
Black Holes and Cosmology'', eds. Unruh W G and Schutzhold R,
Springer Lecture Notes in Physics 718/2007, pp. 31-73.

\bibitem{ruckenstein} M\"oller G,  Ruckenstein A E and Schmitt-Rink S
1992 Transfer of spectral weight in an exactly solvable model of
strongly correlated electrons in infinite dimensions Phys. Rev. B
{\bf 46} 7427-7432.

\bibitem{laughlin}Laughlin R B 1998 A Critique of Two Metals Adv. Phys. {\bf 47} 943-958.

\bibitem{slater} Slater J C 1951 Magnetic Effects and the Hartree-Fock Equation
Phys. Rev. {\bf 82} 538-541.

\bibitem{linio} Kuiper P, Kruizinga G, Ghijsen J, and Sawatzky G A 1989
Character of Holes in Li$_x$Ni${1-x}$O and Their Magnetic Behavior Phys.
Rev. Lett. {\bf 62} 221-224.

\bibitem{p2} Eskes H, Tjeng L H and Sawatzky G A 1990
Cluster-model calculation of the electronic structure of CuO: A
model material for the high-$T_c$ superconductors Phys. Rev. B 41
288-299.

\bibitem{p3} Varma C M, Schmitt-Rink S, Abrahams E 1987
Charge transfer excitations and superconductivity in ionic
metals Solid State Comm. 62 681-685.

\bibitem{p4} Emery V J 1987 Theory of high-$T_c$ superconductivity in oxides
Phys. Rev. Lett. {\bf 58} 2794-2797.

\bibitem{chen} Chen C T, Sette F, Ma Y, Hybertsen M S, Stechel E B, Foulkes W M C, Schulter M, Cheong S W,
Cooper A S, Rupp L W, Batlogg B, Soo Y L, Ming Z H, Krol A and Kao Y
H 1991 Electronic states in La$_{2-x}$Sr$_x$CuO$_{4+\delta}$ probed
by soft-x-ray absorption Phys. Rev. Lett. {\bf 66} 104-107.

\bibitem{sawatzky}Meinders M B J, Eskes H, and Sawatzky G A 1993
Spectral-weight transfer: Breakdown of low-energy-scale sum rules in
correlated systems Phys. Rev. B {\bf 48}, 3916-3926.

\bibitem{stechel2} Hybertsen M S, Stechel E B, Foulkes W M C
and Schl\"uter M 1992 Model for low-energy electronic states probed
by x-ray absorption in high-Tc cuprates Phys. Rev. B {\bf 45},
10032-10050.

\bibitem{leclair1}Kapit E and LeClair A 2008 A unique non-Landau/Fermi liquid in 2d for high Tc superconductivity
arXiv:0805.2951.

\bibitem{leclair2}Tye S H 2008 On the New Model of Non-Fermi Liquid for High Temperature Superconductivity
arXiv:0805.4200.

\bibitem{drew} Stanescu T D, Galitski, and Drew H D, 2008 Effective
  masses in a strongly anisotropic Fermi liquid Phys. Rev. Lett. {\bf
    101} 066405 (2008).

\bibitem{konstanzeros} Stanescu T D and Kotliar G,
2006  Fermi arcs and hidden zeros of the Green function in the
pseudogap state Phys. Rev. B {\bf 74} 125110/1-4.

\bibitem{gauge1}Lee P A, Nagaosa N, Wen, X -G, 2006 Doping a Mott
 insulator: Physics of high-temperature superconductors
 Rev. Mod. Phys. {\bf 78} 17-86. 


\bibitem{gauge2}Anderson P W 2006 The `strange metal' is a projected
 Fermi liquid with edge singularities Nat. Phys. {\bf 3} 626-630.

\bibitem{gauge3} Kotliar G. and
Ruckenstein A E 1986 New functional integral appraoch to strongly
correlated Fermi systems: The Gutzwiller approximation as a saddle point
Phys. Rev. Lett. {\bf 57} 1362-1365.

\bibitem{eskes}Eskes H, Ole\'s A M, Meinders M 
B J, Stephan W, 1994 Spectral properties of the Hubbard bands
Phys. Rev. B {\bf 50}, 17980-18002.

\bibitem{slavery} Phillips P, Galanakis D, and Stanescu T D 2004
Absence of Asymptotic Freedom in Doped Mott Insulators: Breakdown of
Strong Coupling Expansions Phys. Rev. Lett. {\bf 93} 267004.

\bibitem{spalek}Chao K A, Spalek J, and Ole\'s A M 1977 Kinetic exchange interaction in a narrow S-band
J. Phys. C. {\bf 10} L271-L588.

\bibitem{girvin}MacDonald A H, Girvin S M, and Yoshioka D 1988 t/U expansion for the Hubbard model
Phys. Rev. B {\bf 37}, 9753-9756.

\bibitem{sasha} Chernyshev A L, Galanakis D, Phillips P, Rozhkov A V
and Tremblay A M S 2004 Higher order corrections to effective
low-energy theories for strongly correlated electron systems Phys.
Rev. B {\bf 70} 235111/1-12.

\bibitem{anderson} Anderson P W 1959 New Approach to the Theory of Superexchange Interactions
Phys. Rev. {\bf 115} 2-13.

\bibitem{rice}Zhang F C and Rice T M, 1989 Effective Hamiltonian for
 the superconducting Cu oxides Phys. Rev. B {\bf 37}, 3759-3762.

\bibitem{lsu21} Dagotto E, Fradkin E, and Moreo A 1988 SU(2) gauge invariance and order parameters in strongly coupled electronic systems
Phys. Rev. B {\bf  38} 2926-2929.

\bibitem{lsu22} Affleck I, Zou Z, Hsu T, and Anderson P W 1988 SU(2) gauge symmetry of the large-U limit of the Hubbard model
Phys. Rev. B {\bf 38} 745-747.

\bibitem{shole1} Kane C L, Lee P A and Read N 1989 Motion of a single hole in a quantum antiferromagnet
Phys. Rev. B {\bf 39} 6880-6897.

\bibitem{shole2} Dagotto E 1994
Correlated electrons in high-temperature superconductors Rev. Mod.
Phys. {\bf 66} 763-840.

\bibitem{shole22}Mishchenko, A S, Prokof'ev N V and Svistunov B V
  (2001) Single-hole spectral function and spin-charge separation in the t-J model Phys. Rev. B {\bf 64}, 033101/1-4.

\bibitem{shole3} Sorella S 1992 Quantum Monte Carlo study of a single hole in a quantum antiferromagnet
Phys. Rev. B {\bf 46} 11670-11680.

\bibitem{rosch}Haule K, Rosch A, Kroha J, and Wolfle P
2003 Pseudogaps in the t-J model:An extended dynamical mean-field
theory study Phys. Rev. B {\bf 68} 155119/1-19.

\bibitem{prelovsek} Zemljic M M and Prelovsek P
2005  Resistivity and optical conductivity of cuprates within the t-J model
Phys. Rev. B {\bf 72} (2005) 075108/1-8.

\bibitem{holeloc} Choy T P and Phillips P
2005  Doped Mott Insulators Are Insulators: Hole Localization in the
Cuprates Phys. Rev. Lett. {\bf 95} 196405/1-4.

\bibitem{yang2}Kawakami N and Yang S K 1990 Correlation functions in the one-dimensional t-J model Phys. Rev. Lett. {\bf 65}, 2309-2312.

\bibitem{yang} Kawakami N and Yang S K 1990 Luttinger anomaly exponent of momentum distribution in the Hubbard chain Phys. Lett. A {\bf 148}, 359-362.

\bibitem{korepin}Frahm H and Korepin V E 1990 Critical exponents for the one-dimensional Hubbard model Phys. Rev. B {\bf 42}, 10553-10565.

\bibitem{haulek} Haule K and Kotliar G
2007 Optical conductivity and kinetic energy of the superconducting
state: A cluster dynamical mean field study Europhys. Lett. {\bf 77}
27007-27010.

\bibitem{zaanen} Mischenko A S, Nagaosa N, Shen Z X, De Filippis G,
Cauaudella V, Devereaux T P, Bernhard C, Kim K W, and Zaanen J 2008
Charge Dynamics of Doped Holes in High Tc Cuprate Superconductors: A Clue from Optical Conductivity
Phys. Rev. Lett. {\bf 100} 166401/1-4.

\bibitem{millis} Comanac A, de Medici L, Capone M, and Millis A J
2008 Optical conductivity and the correlation strength of
high-temperature copper-oxide superconductors Nat. Phys. {\bf 4}
287-290.

\bibitem{chakrab} Chakraborty S, Galanakis D, and Phillips P
2008 Kinks and Mid-Infrared Optical Conductivity from Strong
Electron Correlation arXiv:0712.2838.

\bibitem{dzy} Dzyaloshinskii I 2003
Some consequences of the Luttinger theorem: The Luttinger surfaces in
non-Fermi liquids and Mott insulators Phys. Rev. B {\bf 68},
85113/1-6.

\bibitem{zeros} Phillips P 2006 Mottness Ann. of Phys. {\bf 321} 1634-1650.

\bibitem{zeros2}
Stanescu T D, Phillips P, and Choy T P 2007  Theory of the Luttinger
surface in doped Mott insulators Phys. Rev. B {\bf 75} 104503/1-9.

\bibitem{essler} Essler F H L and Tsvelik A,
2003 Finite Temperature Spectral Function of Mott Insulators and
Charge Density Wave States Phys. Rev. Lett. {\bf 90}126401/1-4

\bibitem{bohm} Bohm D and Pines D
1953 A Collective Description of Electron Interactions: III. Coulomb
Interactions in a Degenerate Electron Gas Phys. Rev. {\bf 92}
609-625.

\bibitem{sm}Murthy G and Shankar R, 2003 Hamiltonian theories of the
 fractional quantum Hall effect Rev. Mod. Phys. {\bf 75}, 1101-1158.

\bibitem{cluster1} Moukouri S and Jarrell M,
2001 Absence of a Slater Transition in the Two-Dimensional Hubbard
Model Phys. Rev. Lett. {\bf 87} 167010/1-4.

\bibitem{cluster2} Civelli M, Capone M, Kancharla S S, Parcolet O,
and Kotliar G 2005  Dynamical Breakup of the Fermi Surface in a
Doped Mott Insulator Phys. Rev. Lett. {bf 95} 106402/1-4.

\bibitem{mg1} Maier T, Jarrell M, Pruschke T, and Hettler M H 2005
  Rev. Mod. Phys. {\bf 77} 1027-1081.

\bibitem{unpub} Leigh R G and Phillips P Origin of the Mott gap,
  arXiv:0812.0593.

\bibitem{lenkink} Lanzara A, Bogdanov P V, Zhou X J, Kellar S A, Feng D L, Lu E D,
Yoshida T, Eisaki H, Fujimori A, Kishio K, Shimoyama J I, Noda T,
Uchida S, Hussain Z and Shen Z X 2001  Evidence for ubiquitous
strong electron-phonon coupling in high-temperature superconductors
Nature {\bf 412} 510-514.

\bibitem{henkink} Graf J, Gweon G H, McElroy K, Zhou D Y, Jozwiak C, Rotenberg E, Bill A, Sasagawa T,
Eisaki H, Uchida S, Takagi H, Lee D H, and Lanzara1 A 2007 Universal
High Energy Anomaly in the Angle-Resolved Photoemission Spectra of
High Temperature Superconductors: Possible Evidence of Spinon and
Holon Branches Phys. Rev. Lett. {\bf 98} 67004.

\bibitem{jkink} Macridin A, Jarrell, M, Maier, T A, and Scalapino D J (2007)
  High energy kink in the single particle spectra of the
  two-dimensional Hubbard model Phys. Rev. Lett.  {\bf 99}
  237001/1-4.

\bibitem{pink}Zemljic M M, Prelovsek P, and Tohyama T (2007)
  Temperature and doping dependence of high-energy kink in cuprates
  Phys. Rev. Lett. {\bf 100}, 036402/1-4.

\bibitem{nopg}Fisher D S, Kotliar G and Moeller, G 1995 Midgap
  states in doped Mott insulators in infinite dimensions, Phys. Rev. B
  {\bf 52}, 17112-17118.

\bibitem{pg1} Maier T, Jarrell M, Pruschke T and Hettler M H, 1027
  Quantum Cluster Theories {\bf 77}, 1028-1090.

\bibitem{pg2}Stanescu T D and Phillips P 2003 Pseudogap in doped Mott
  insulators is the Near-neighbour analogue of the Mott gap,
  Phys. Rev. Lett. {\bf 91}, 017002/1-4.

\bibitem{int1} Hwang J, Timusk T and Gu G D,
2007  Doping dependent optical properties of
Bi$_2$Sr$_2$CaCu$_2$O$_{8+\delta}$ J. Phys. Condens. Matter {\bf 19}
125208-125240.

\bibitem{int2}Onose Y, Taguchi Y, Ishizaka K, and Tokura Y
2004 Charge dynamics in underdoped Nd$_{2-x}$Ce$_x$CuO$_4$: Pseudogap
and related phenomena Phys. Rev. B {\bf 69} 024504/1-13.

\bibitem{int3}Lucarelli A, Lupi S, Ortolani M, Calvani P, Maselli P, Capizzi M,
Giura P, Eisaki H, Kikugawa N, Fujita T, Fujita M, and Yamada K 2003
Phase Diagram of La$_{2-x}$Sr$_x$CuO$_4$ Probed in the Infared:
Imprints of Charge Stripe Excitations Phys. Rev. Lett. {\bf 90}
037002/1-4.

\bibitem{zaanen2} Cvetkovic V, Nussinov Z, Mukhin S, and Zaanen J,
2007 Observing the fluctuating stripes in high-Tc superconductors
Europhys. Lett. {\bf 81} 27001-27004.

\bibitem{loram} Loram J W, Luo J, Cooper J R, Liang W Y and Tallon J L,
2001 Evidence on the pseudogap and condensate from the electronic
specific heat J. Phys. and Chem. Sol. {\bf 62} 59-64.

\bibitem{tlin} Phillips P and Chamon C
2005  Breakdown of One-Parameter Scaling in Quantum Critical
Scenarios for High-Temperature Copper-Oxide Superconductors Phys.
Rev. Lett. {\bf 95} 107002/1-4.

\bibitem{bass} Bass J, Pratt W P, and Schroeder P A,
1990 The temperature-dependent electrical resistivities of the
alkali metals Rev. Mod. Phys. {\bf 62} 645-744.

\bibitem{vanishingtp}Honma T and Hor P H Unified electronic phase
  diagram for hole-doped high-$T_c$ cuprates 2008 Phys. Rev.
B {\bf 77}, 184520/1-16.

\bibitem{thermoboys} Presland J, Tallon J L, Buckley R G, Liu R S and
  Flower N E, 1991 General trends in oxygen stochiometry effects on
  $T_c$ in Bi and Tl supercoductors, Physica C {\bf 176} 95-105.

\bibitem{beni}Beni G, 1973 Thermoelectric power of the narrow-band
 Hubbard chain at arbitrary electron density: Atomic limit Phys. Rev B {\bf 10} 2186-2189.

\bibitem{shila} Chakraborty S, Galanakis D and Phillips P Emergence of
  particle-hole symmetry near optimal doping in the high-temperature
  copper-oxide superconductors,
  arXiv:0807.2854.

\bibitem{mkhe} R\"ubhausen M, Gozar A, Klein M V, Guptasarma P,
and Hinks D G, 2001 Superconductivity-induced optical changes for energies
of $100\Delta$ in the cuprates Phys. Rev. B {\bf 63}, 224514/1-5.


\bibitem{bontempshe}Santander-Syro A F,  Lobo R P S M,  Bontemps N,
 Konstantinovic Z, Li Z Z, and Raffy H 2003  Pairing in cuprates from high energy electronic states Europhys. Lett. {\bf 62}, 568-574.

\bibitem{vmhe}Molegraaf H J A,  Presura C,   van der Marel D, 
Kes P H, and Li M, 2002 Superconductivity-induced transfer of in-plane spectral weight
in Bi$_ 2$Sr$_2$CaCu$_ 2$O$_{8+ \delta}$ Science, {\bf 295} 2239-2241.

\end{thebibliography}
\end{document}